\documentclass[11pt,a4paper]{article}
\usepackage{jheppub,xcolor}
\usepackage[utf8]{inputenc}
\usepackage{textcomp}
\usepackage{newunicodechar}
\newunicodechar{γ}{\ensuremath{\gamma}}
\usepackage{blindtext}
\usepackage{physics}
\usepackage{caption}
\usepackage{booktabs}
\usepackage{subcaption}
\usepackage{slashed}
\usepackage{bigints}
\usepackage{setspace}
\usepackage{array}
\usepackage{comment}
\usepackage{multirow}
\usepackage{makecell} 
\definecolor{mygreen}{rgb}{0, 0.5019607843137255, 0}
\definecolor{Red}{rgb}{1.,0.,0.}
\definecolor{Blue}{rgb}{0.,0.,1.}
\usepackage{pstricks}
\usepackage{xcolor}
\usepackage{amsmath,amsfonts,amsthm,amssymb,bm,bbm}
\usepackage{epsf}
\usepackage{epsfig}
\usepackage{afterpage}
\usepackage{longtable}
\usepackage{latexsym,mathrsfs,dsfont}
\usepackage{graphics}
\usepackage{url}
\usepackage{paralist}
\usepackage{bbold}
\usepackage[colorlinks=true, urlcolor=blue]{hyperref}
\newcommand{\nn}{\nonumber}

\newcommand{\spur}[1]{\not\! #1 \,}
\newcommand{\be}{\begin{equation}}
\newcommand{\ee}{\end{equation}}
\newcommand{\bea}{\begin{eqnarray}}
\newcommand{\eea}{\end{eqnarray}}

\newcommand{\DD}{\stackrel{\leftrightarrow}{\partial}}
\newcommand{\DDleft}{\stackrel{\leftarrow}{\partial}}
\newcommand{\DDright}{\stackrel{\rightarrow}{\partial}}
\newcommand{\DDcov}{\stackrel{\leftrightarrow}{D}}
\newcommand{\DDcovleft}{\stackrel{\leftarrow}{D}}
\newcommand{\DDcovright}{\stackrel{\rightarrow}{D}}
\newcommand{\mup}{\mu^\prime}
\newcommand{\nup}{\nu^\prime}
\newcommand{\mHsq}{m_{H_2^*}}
\newcommand{\sigmabar}{\bar \sigma}

\allowdisplaybreaks

\title{Semileptonic $B_q$  decays to heavy tensor mesons}

\author[a]{Pietro~Colangelo,}
\author[a]{Fulvia~De~Fazio,}
\author[b]{Carlo~la~Torre,}
\author[a,b]{and Giuseppe~Roselli\,}

\emailAdd{pietro.colangelo@ba.infn.it} \emailAdd{fulvia.defazio@ba.infn.it} \emailAdd{c.latorre12@alumni.uniba.it} \emailAdd{giuseppe.roselli@ba.infn.it}

\affiliation[a]{Istituto Nazionale di Fisica Nucleare, Sezione di Bari,  Via Orabona 4, 70126 Bari, Italy}
\affiliation[b]{ Dipartimento Interateneo di Fisica "Michelangelo Merlin", Universit\`a degli Studi di Bari, via Orabona 4, 70126 Bari, Italy}

\abstract{The semileptonic decays of $B_q$ mesons ($q=u,d,s$) to $J^P=2^+$ charmed mesons are investigated. The  form factors parametrising the hadronic matrix elements of the weak vector and  axial-vector  quark currents  are evaluated using  light-cone QCD sum rules with external $B_q$ state, including the 3-particle contributions. The form factors of  pseudoscalar and tensor quark currents occurring in extensions of the Standard Model are also determined. The relations expected in the  heavy quark  limit are tested,  providing a hint about   the size of finite heavy quark mass corrections  for the various form factors. Results are presented for the semileptonic $B_q$ to heavy  tensor meson decay rates in the Standard Model, and for the ratios testing  Lepton Flavour Universality.}

\begin{document} 

\begin{flushright}  
BARI-TH/26-783
\end{flushright}

\maketitle

\section{Introduction}

Semileptonic decays of $B_q$ mesons ($q=u,d,s$) to charmed mesons with spin-parity $J^P=2^+$ are significant for many reasons. 
These processes, induced at the quark level by the  $b \to c \ell \bar\nu_\ell$ transition, provide additional methods for determining the Cabibbo-Kobayashi-Maskawa (CKM) matrix element $|V_{cb}|$,   a  fundamental parameter of the Standard Model (SM) \cite{Gambino:2020jvv}. Currently,  measurements of this parameter using exclusive channels are  dominated by   $\bar B \to D^{(*)} \ell \bar\nu_\ell$ modes and their $\bar B_s$ counterparts \cite{HeavyFlavorAveragingGroupHFLAV:2024ctg}. However, a sizable fraction of the inclusive $\bar B \to X_c \ell \bar\nu_\ell$
rate consists of high-mass charmed mesons, such as  orbital excitations.  A precise understanding  of this contribution is mandatory to  complete
the exclusive description of  semileptonic $B_q$ modes and to  reduce    systematic uncertainties in   $|V_{cb}|$ determinations.

Furthermore,   $B_q$ decays to heavy tensor mesons can serve as a probe for New Physics (NP).  They allow for the scrutiny of potential  deviations from SM predictions, which would signal the presence of  operators not included in the SM weak effective Hamiltonian or modifications of the SM Wilson coefficients. Various observables, such as the angular distributions of the final mesons,  can be constructed and compared with 
 experimental data or correlated with  distributions for the $B_q \to D^{(*)}_q$ modes \cite{Colangelo:2018cnj,Colangelo:2024mxe}. Additionally, these decays allow for  tests of  the lepton flavour universality (LFU) - an accidental symmetry of SM \cite{Biancofiore:2013ki,Colangelo:2016ymy,Bernlochner:2021vlv}. This complements the information obtained  from   $D^{(*)}_q$ channels, where  persistent tensions have been observed \cite{HeavyFlavorAveragingGroupHFLAV:2024ctg}.  

From the  QCD perspective, these decay modes are important to resolve the so-called $1/2-3/2$ puzzle \cite{Bigi:2007qp}. This issue concerns the weak form factors involved in $B_q$ transitions to the  lightest four positive-parity charmed mesons.  In the heavy quark limit ($m_Q \to \infty$), the four $Q \bar q$ states with $J^P=0^+, 1^+, 2^+$ can be organized in two mass-degenerate spin doublets categorized by the  total angular momentum  of the light degrees of freedom $j_\ell = 1/2$ and $j_\ell = 3/2$.  All form factors  for these transitions
can  be expressed in terms of two universal functions, $\tau_{1/2}$ and $\tau_{3/2}$,  analogous to the Isgur-Wise function 
\cite{Isgur:1990yhj,Shifman:1987rj,Falk:1990yz,Neubert:1993mb}. Unlike the Isgur-Wise function, however, 
 $\tau_{1/2}$ and $\tau_{3/2}$  are not determined by the heavy quark symmetry at the zero-recoil point,  the maximal momentum transfer to the lepton pair. Consequently, their values and the magnitude
 of the finite heavy quark mass corrections  for the form factors remain subject of debate \cite{Colangelo:1998ga,Colangelo:2000jq,LeYaouanc:2014pco,Wang:2022tuy}.
Assessing the limitations of current QCD-based computational tools is essential for  determining these hadronic matrix elements.

On the experimental side, a $J^P = 2^+$ state such as  $D_2^*(2460)$  (falling into the 
$j_\ell = 3/2$ spin doublet)  is favourable to study due to the narrow full width and the possibility to set up several measurable observables in semileptonic $B$ decays. However, precision must be improved in the  determination of the form factors parametrizing the hadronic matrix elements of quark currents, both  in  SM and in its  extensions, in the full kinematical range.
While there are progresses in lattice QCD  calculation of the $\bar B \to D^{(*)}$ form factors, the calculations involving excited and unstable final-state mesons remain challenging.
An issue is that  the  resonant $D^{(*)}\pi$ states make difficult  the extraction of matrix elements associated to tensor resonances. Moreover, the range of squared momentum transfer reliably accessible to computation  is limited to the kinematics near to  zero-recoil \cite{Atoui:2013ksa}. A different QCD-based  approach,  light-cone QCD sum rules (LCSRs) \cite{Balitsky:1986st,Balitsky:1989ry,Chernyak:1990ag}
 \footnote{The method is described in detail in  refs.\cite{Colangelo:2000dp,Khodjamirian:2020btr}.},  is able to reliably access the complementary kinematical range, hence a possible integration of  the two methods can be foreseen. In particular, the version of LCSRs employing  $B_q$-meson light-cone distribution amplitudes (LCDA)  has proven to be a powerful and flexible framework for the calculation of 
form factors at large hadronic recoil \cite{DeFazio:2005dx,Khodjamirian:2005ea,Khodjamirian:2006st,DeFazio:2007hw}.
In this approach, the nonperturbative QCD dynamics is encoded in universal $B_q$-meson light-cone
distribution amplitudes, and the quantum numbers of the final meson are selected
through appropriate interpolating currents in correlation functions,  the basic quantities to be scrutinized.
This version of  LCSRs has been applied to   $\bar B_{u,d} \to D_{u,d}^{(*)}$  and to related $B_s$ transitions (as reviewed  in \cite{Khodjamirian:2023wol}), as well as to   $B_{u,d}$  decays to $J^P=0^+,1^+$ charmed mesons 
 \cite{Gubernari:2022hrq,Gubernari:2023rfu}.

In this work we apply light-cone QCD sum rules  with $B$ LCDAs to the calculation of the  $\bar B_q \to D_{q2}^*$ form factors
keeping the light-quark mass.
In the  sum rules  we include the contributions from two-particle  and three-particle $B$-meson LCDAs up  to twist-four accuracy,  while higher-twist contributions, such as those associated with $g_-$  and the three-particle LCDA $\phi_6$, are included only through Wandzura–Wilczek type relations and equations of motion constraints.
The resulting form factors are used to provide predictions for the differential and integrated decay widths of the modes $\bar B_q \to D_{q2}^* \ell \bar\nu_\ell$.
We compare the results with the expectations based on heavy-quark symmetry in the large $m_Q$ limit,  and with other existing calculations. We also discuss the phenomenological implications for the  ongoing 
and future measurements.

The paper is organized as follows.
In section~\ref{sec:twopt} we describe the two-point QCD sum rule calculation of the constant $f_{H^*_2}$ parametrizing the vacuum-particle matrix element of a current interpolating the heavy $Q \bar q$ tensor mesons  $H^*_2=D^*_{(s)2}$ and $B^*_{(s)2}$.
Section~\ref{sec:LCSR} is devoted to the derivation of light-cone QCD sum rules for the
$ B_q \to D_{q2}^*$ form factors, with numerical  results  presented in sec.~\ref{sec:Numerics}.
 Phenomenological implications  are discussed  in section~\ref{sec:pheno}, together with a comparison with the results of previous studies.
The conclusions and the outlook for future improvements are summarized in section~\ref{sec:Conclusion}.
The appendices comprise details relevant for the analysis.
Appendix~\ref{appA} contains the expressions of the perturbative spectral function and of the contribution from the D=3,4,5 operators  in the two-point QCD sum rule calculation of  $f_{H^*_2}$.
In appendix~\ref{appB} the formulae relevant for the LCSR calculation of the $B_q \to D^*_{q2}$ form factors are collected.
Appendix~\ref{appC} comprises the expressions and  the parameters of the $B_q$  LCDA  used in the form factor determination.

\section{Decay constant $f_{H^*_2}$ of the $Q \bar q$ tensor meson}\label{sec:twopt}
\subsection{Two-point QCD sum rule for $f_{H^*_2}$}
For the calculation of the form factors parametrizing the $B_q \to D_{q 2}^*$ weak matrix elements using light-cone QCD sum rules  we need  the  decay constant of $D_{q 2}^*$.
We  generically  denote as $H_2^*$ a $Q \bar q$ meson with $J^P=2^+$,   because the calculation also holds for the  beauty mesons $B_{q2}^*$. 
The decay constant  $f_{H_2^*}$  is defined  as
\be
\langle 0|J_{\mu \nu}|H_2^*(p,\lambda) \rangle=m_{H_2^*}^2\, f_{H_2^*} \,  \epsilon_{\mu \nu}(p, \lambda)  \,\,\, , \label{fTdef}
\ee
where $m_{H_2^*}$ and $\epsilon_{\mu \nu}(p,\lambda)$  are the mass and the polarization tensor of $H_2^*$.  In \eqref{fTdef}  $J_{\mu \nu}$ is a quark current interpolating a  meson  with $J^P=2^+$. 
We use the symmetric and traceless current

\be
J_{\mu \nu}(x)=i \,{\bar q}(x) \left(\DDcov_\mu \gamma_\nu +\DDcov_\nu \gamma_\mu-\frac{1}{2}g_{\mu \nu}\spur \DDcov \right) Q(x) =i\, A_{\mu \nu}^{\alpha \beta} \,\,  {\bar q}(x)\DDcov_\alpha \gamma_\beta Q(x) \,\, ,  \label{current}
\ee
 first proposed  for the application of QCD sum rules to tensor mesons  \cite{Aliev:1981ju}.
 $Q=c,b$ and   $q=u,d,s$  are the heavy and light quark fields, respectively.  We define  $\DDcov_\mu=\DDcovright_\mu-\DDcovleft_\mu$, with   covariant derivatives  
 $\displaystyle\DDcovright_\mu=\DDright_\mu -i g_s \frac{\lambda^a}{2} G^a_\mu$ and
$\displaystyle \DDcovleft_\mu=\DDleft_\mu + i g_s \frac{\lambda^a}{2} G^a_\mu$. $G$ is the gluon field.
  The factor $A_{\mu \nu}^{\alpha \beta}$ in \eqref{current} is defined as    $A_{\mu \nu}^{\alpha \beta}=g_\mu^\alpha g_\nu^\beta+g_\nu^\alpha g_\mu^\beta-\displaystyle\frac{1}{2}g_{\mu \nu} g^{\alpha \beta}$.

We consider the two-point correlator \cite{Shifman:1978bx}
\be
\Pi_{\mu \nu \mu^\prime \nu^\prime}(q)=i\, \int d^4x e^{i q \cdot x} \langle 0|T\left\{J_{\mu \nu}(x) J_{\mu^\prime \nu^\prime}^\dagger(0) \right\} |0 \rangle \,\, . \label{2ptcorr}
\ee
To  single out the contribution of the $J^P=2^+$ state we introduce
\bea
\pi_{\alpha \beta}(q)&=&-g_{\alpha \beta}+\frac{{q_\alpha}q_{\beta}}{q^2} \,\, \\
{\cal P}_{\mu \nu \mup \nup}(q)&=&\frac{1}{2}\left(\pi_{\mu \mup}(q) \pi_{\nu \nup}(q)+\pi_{\mu \nup}(q) \pi_{\nu \mup}(q)-\frac{2}{3}\pi_{\mu \nu}(q) \pi_{\mup \nup}(q)\right)
\label{projector}
\eea
and  we consider  the scalar function obtained by the product
\be
\Pi(q^2)={\cal P}^{\mu \nu \mup \nup}(q)\Pi_{\mu \nu \mu^\prime \nu^\prime}(q)  \,\, . \label{2ptcontracted}
\ee
A pedagogical description of the QCD sum rule analysis of a two-point correlator can be found in \cite{Colangelo:2000dp}. Here we briefly outline  the calculation.
 $\Pi(q^2)$  satisfies a subtracted dispersion relation
\be
\Pi(q^2)=\frac{1}{\pi}\int_{s_{\rm min}}^\infty ds \, \frac{{\rm Im }\,\Pi(s)}{s-q^2-i \epsilon}+ {\rm subtractions}\,\,, \label{twop-dispersion}
\ee
the subtraction terms will  be dealt with at the end.
Using unitarity,  ${\rm Im }\,\Pi(s)$ is obtained inserting in \eqref{2ptcorr} a complete set of  states coupled to the current $J_{\mu \nu}$. The lightest narrow one-particle state $|H_2^*\rangle$  is singled out,  all other contributions being included in the hadronic  spectral function  $\rho_h$  for $s$   above a threshold $s_0$.   This is the  one-particle + continuum model for the spectral density:
 \be
 \frac{1}{\pi}{\rm Im }\,\Pi(s)={\cal P}^{\mu \nu \mup \nup} \langle 0|J_{\mu \nu}|H_2^* \rangle \langle H_2^*|J_{\mu^\prime \nu^\prime}^\dagger|0\rangle \delta(s-m_{H_2^*}^2)+\rho_h(s) \theta(s-s_0) \,\,.
 \ee
 The identification allows to express  $\Pi(q^2)$ in terms of hadronic quantities
 \be
\Pi^{\rm had}(q^2)=5\frac{m_{H_2^*}^4 \, f_{H_2^*}^2}{m_{H_2^*}^2-q^2} + \int_{s_0}^\infty ds \, \frac{\rho_h(s)}{s-q^2} + {\rm subtractions}\,\,. \label{hadronic}
\ee

On the other hand, for large $Q^2=-q^2$  the correlator  \eqref{2ptcorr} is dominated by short-distances. By an operator product expansion (OPE) it can be expressed as the sum of a perturbative contribution plus nonperturbative terms proportional to quark and gluon vacuum condensates:
 \be
 \Pi^{\rm QCD}(q^2)=\Pi^{\rm pert}(q^2)+\Pi^{\rm non-pert}(q^2)\,\,. \label{QCD-generic}
 \ee
 The perturbative term is  written  (modulo subtractions) as
 \be
\Pi^{\rm pert}(q^2)=\int_{(m_Q+m_q)^2}^\infty ds \, \frac{\rho^{\rm pert}(s)}{s-q^2-i \epsilon} \, .
\ee
The expression of the perturbative spectral density $\rho^{\rm pert}(s)$  at LO in QCD is  in appendix \ref{appA}.  NLO corrections require a two-loop calculation beyond the purposes of the present analysis.
As for  the nonperturbative terms, we truncate the OPE considering only the terms proportional to the quark vacuum condensate, the gluon condensate and the mixed quark-gluon condensate, resulting from operators of dimension $D=3,\,4,\,5$. The term proportional to the gluon vacuum condensate has an imaginary part for $q^2>(m_Q+m_q)^2$ and can be represented by a dispersion relation.  We treat it together with  the perturbative term,  defining $\rho^{\rm QCD}(s)=\rho^{\rm pert}(s)+\rho^{D=4}(s)$, with  $\rho^{D=4}(s)$  in appendix \ref{appA}.

A comment is in order concerning  mesons with strangeness, with  $m_s \neq 0$. 
In this case the imaginary part of the coefficient of the gluon condensate develops a nonintegrable 
threshold singularity when integrated over the  range 
$[(m_Q+m_s)^2,\,s_0]$. This can be inferred from eqs.~\eqref{g21}–\eqref{g25}, finding  that
 near  $s_{\rm th}=(m_Q+m_q)^2$ the spectral density behaves as
$\rho^{(D=4)}(s)\sim (s-s_{\rm th})^{-3/2}$,
reflecting the nonuniform validity of the local OPE near the physical threshold. 
 This was already noticed in QCD sum rules \cite{Bagan:1994dy}.
We treat this contribution in the sense of generalized distributions, 
by a prescription  analogous to the plus–prescription \cite{Peskin:1995ev}.
Anticipating the Borel transformation which is defined below, we write
\be
\widehat \Pi^{(D=4)}(M^2)=
\int_{s_{\rm th}}^{s_0} ds\, \rho^{(D=4)}(s)\, e^{-s/M^2} = \int_{s_{\rm th}}^{s_0} ds\, \frac{G(s)}{(s-s_{\rm th})^{3/2}} \,\,\, ,\label{G2-bor}
\ee
where
\be
G(s)=\rho^{(D=4)}(s)\,(s-s_{\rm th})^{3/2} e^{-s/M^2} 
\ee
 is regular for $s\to s_{\rm th}$. The integral is defined via the subtraction
\be
\int_{s_{\rm th}}^{s_0} ds\, \frac{G(s)}{(s-s_{\rm th})^{3/2}} \equiv
\int_{s_{\rm th}}^{s_0} ds\, \frac{G(s)-G(s_{\rm th})}{(s-s_{\rm th})^{3/2}} -2\,\frac{G(s_{\rm th})}{\sqrt{s_0-s_{\rm th}}}\, ,
\ee
which follows from integrating the singular term in a distributional sense:
$\int dx\, x^{-3/2} = -2 x^{-1/2}$.
With this prescription the result is finite for nonvanishing light–quark mass, 
as in the strange case, and smoothly connects to the $m_q\to0$ limit.
Analogous prescriptions based on generalized ($*$) distributions are adopted 
 to treat similar endpoint singularities \cite{DeFazio:1999ptt, Moreno:2022goo}.~\footnote{ The regularization procedure  adopted in \cite{Bagan:1994dy} consists of computing  $\tilde G(s)$ from the expansion of  $G(s)$ in powers of $s-s_{th}$ up to the first order, and  replacing  $G(s)$ with $G(s)-\tilde G(s)$ in eq.~\eqref{G2-bor}.}
 
The last step of the calculation consists in  invoking semiglobal quark-hadron duality:  we replace \footnote{ Quark-hadron duality usually amounts to identifying  integrals of the hadronic spectral function  $\rho_h(s)$  with integrals of the perturbative spectral function $\rho^{\rm pert}(s)$.  We also include  the OPE D=4 contribution in the latter one.} 
\be
\rho_h(s) \to \rho^{\rm QCD}(s) \,\,\, 
\ee
 in the dispersive integral above the continuum threshold $s_0$. We obtain (modulo subtractions)
\be
5\frac{m_{H_2^*}^4 \, f_{H_2^*}^2}{m_{H_2^*}^2-q^2}=\int_{(m_Q+m_q)^2}^{s_0} ds \, \frac{\rho^{\rm QCD}(s)}{s-q^2-i \epsilon}+\Pi^{D=3}(q^2) +\Pi^{D=5}(q^2) \label{sr} \,\,.
\ee
To suppress the contribution of heavier resonances and of the continuum, a Borel transform with respect to $Q^2$  is applied, which  also allows us  to get rid of the subtraction terms, polynomials in $q^2$.
Using 
\be
{\widehat B}\frac{1}{(m^2+Q^2)^n}=\frac{1}{(n-1)!}\frac{e^{-m^2/M^2}}{(M^2)^{(n-1)} }     \label{borel} 
\ee
 the  sum rule follows: 
\be
5m_{H_2^*}^4 \, f_{H_2^*}^2\,e^{-\mHsq/M^2}=\int_{(m_Q+m_q)^2}^{s_0} ds \,  \rho^{\rm QCD}(s)\,e^{-s/M^2}+{\widehat \Pi}^{D=3}(M^2) +{\widehat \Pi}^{D=5}(M^2)\label{srborel} \,\,.
\ee
The hat  indicates  Borel transformed terms.
The expressions of ${\widehat \Pi}^{D=3}$ and ${\widehat \Pi}^{D=5}$ are   in appendix \ref{appA}.

The continuum threshold $s_0$ and the Borel variable $M^2$  in \eqref{srborel} are parameters. Their range can be determined
 considering  the derivative of \eqref{srborel} with respect to $\displaystyle -\frac{1}{M^2}$. 
In the ratio between the  resulting sum rule and the original one, the   constant $f_{H^*_2}$ drops out and  an equation for the mass $m_{H_2^*}$ is obtained. The  measured mass can be used to set  the  parameters in the various cases, as we discuss in the following.

\subsection{Numerics for $f_{H^*_2}$}

In  eq.~\eqref{srborel} the input quantities are  the quark masses and  the  vacuum condensates  in the right-hand QCD side of the sum rule, and the meson masses in the left-hand
hadronic side. For the quark masses  we use $m_c(m_c)=1.273 \pm 0.005~\mathrm{GeV}$,
$m_b(m_b)=4.183 \pm 0.007~\mathrm{GeV}$ and $m_s(2~\mathrm{GeV})=93^{+11}_{-5}\ \mathrm{MeV}$  in the $\overline{\mathrm{MS}}$ scheme. We neglect the mass of up and down quarks.
The  values of the QCD vacuum condensates are  in table~\ref{tab:cond}.
\begin{table}[t]
\centering
\renewcommand{\arraystretch}{1.3}
\begin{tabular}{lcc}
\hline \hline
 & $\mu = 1.3~\mathrm{GeV}$ &  $\mu = 4.2~\mathrm{GeV}$ \\
\hline
$\langle \bar{q} q \rangle(\mu)$ & $-\,(0.285~\mathrm{GeV})^3$ & $-\,(0.254~\mathrm{GeV})^3$ \\
$\langle \bar{s} s \rangle(\mu)$ & $-\,(0.260~\mathrm{GeV})^3$ & $-\,(0.231~\mathrm{GeV})^3$ \\
$\left\langle \dfrac{\alpha_s}{\pi} G^2 \right\rangle$ & $0.012 ~\mathrm{GeV}^4$ & $0.012~\mathrm{GeV}^4$ \\
$m_{0,q}^2=\dfrac{\langle\bar{q}\,g_s\sigma G\,q\rangle(\mu)}{\langle\bar{q}q\rangle(\mu)}$ &  
$\;0.82~\mathrm{GeV}^2 $ &
$\;0.75~\mathrm{GeV}^2 $ \\
$m_{0,s}^2=\dfrac{\langle\bar{s}\,g_s\sigma G\,s\rangle(\mu)}{\langle\bar{s}s\rangle(\mu)}$ & 
$\;0.80~\mathrm{GeV}^2 $ &
$\;0.73~\mathrm{GeV}^2 $ \\
\hline \hline
\end{tabular}
\caption{\baselineskip 10pt \small Condensates at the charm and beauty scale in the $\overline{\mathrm{MS}}$ scheme.}\label{tab:cond}
\end{table}
The measured meson masses are quoted in table~\ref{tab:par}.
\begin{table}[b!]
\centering
\renewcommand{\arraystretch}{1.3}
\begin{tabular}{lccccc}
\hline \hline
$H_2^*$ \hskip 0.1cm & $m^{exp}_{H^*_2} \hskip 0.1cm  [GeV]$  & $\Gamma^{exp}(H^*_2) \hskip 0.1cm  [MeV]$ &
$s_0 \hskip 0.1cm  [GeV^2] $ \hskip 0.1cm & $M^2\hskip 0.1cm  [GeV^2]$\hskip 0.1cm & $ f_{H^*_2} \hskip 0.1cm [GeV]$\\
\hline
$D_2^*$ &$2.4611\pm 0.0008$ 
& $47.3 \pm 0.8$ &
 $9.2\pm 0.4$ & $3.1 \,-\,  3.6 $ &  $ 0.264 \pm 0.011 \pm 0.017$\\
$D_{s2}^*$ &$ 2.5691\pm 0.0008$ & 
$16.9 \pm 0.7 $
&$9.6\pm 0.4$ & $2.9\,-\, 3.8 $ & $ 0.274 \pm 0.012  \pm 0.017$\\
$B_{2}^*$ &$  5.7396 \pm 0.0007$  &
$20 \pm 5$
&$40.2\pm 0.8$ & $7.0 \, - \, 8.8$ & $0.172 \pm 0.006 \pm 0.036$\\
$B_{s2}^*$ &$5.83988 \pm 0.00012 $ &
$1.49 \pm 0.27$
&$41.0\pm 0.8$ & $7.0\, - \, 9.2$ & $ 0.188 \pm 0.007 \pm 0.039$\\
\hline \hline
\end{tabular}
\caption{\baselineskip 10pt \small Measured  masses  and widths of the heavy tensor mesons $H_2^*$ {\color{blue} \cite{ParticleDataGroup:2024cfk},  }ranges of the parameters $s_0$ and $M^2$ in the related sum rules, and  results for the decay constant $ f_{H^*_2}$. The first error in the last column  reflects the variation of the Borel parameter and continuum threshold,  the second error is obtained from the variation of the renormalization scale entering the running heavy-quark masses.}\label{tab:par}
\end{table}

\begin{figure}[t]
\begin{center}
\includegraphics[width = 0.48\textwidth]{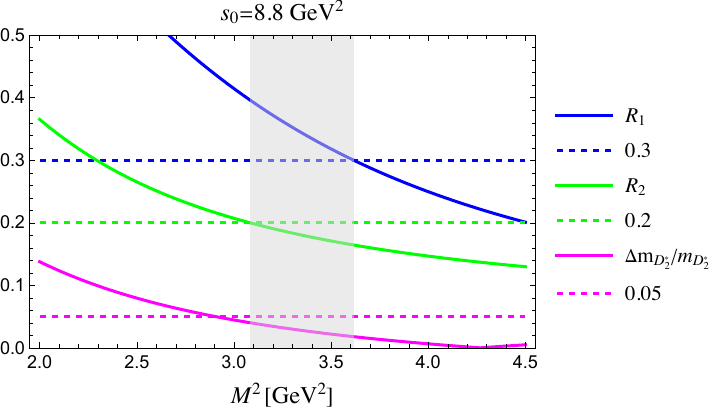} \hskip 0.2cm
\includegraphics[width = 0.48\textwidth]{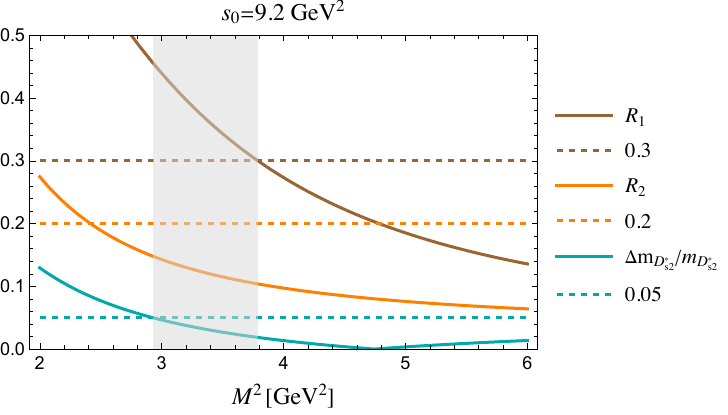}\\
\vskip 0.3cm
\includegraphics[width = 0.48\textwidth]{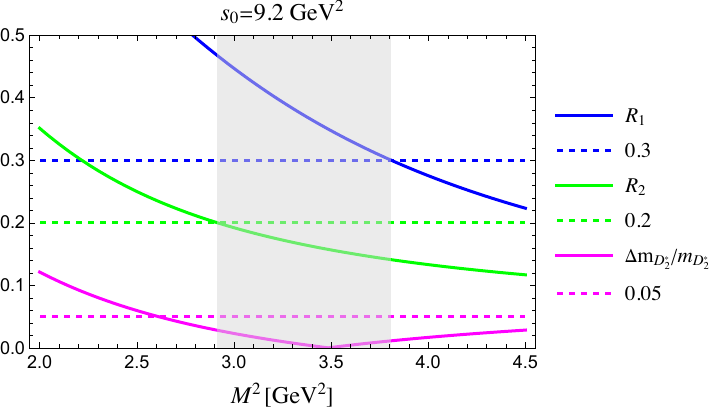} \hskip 0.2cm
\includegraphics[width = 0.48\textwidth]{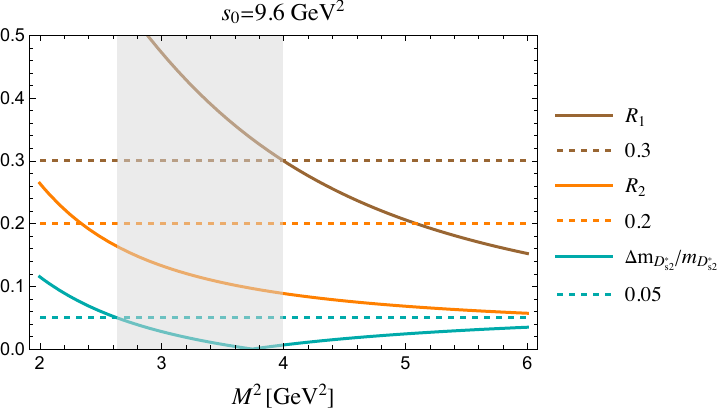}\\
\vskip 0.3cm
\includegraphics[width = 0.48\textwidth]{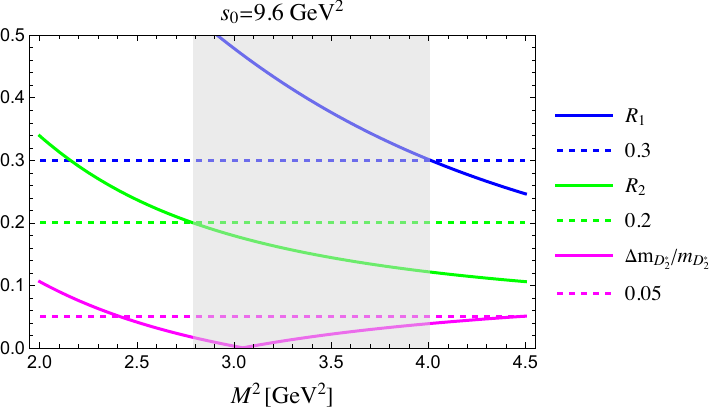} \hskip 0.2cm
\includegraphics[width = 0.48\textwidth]{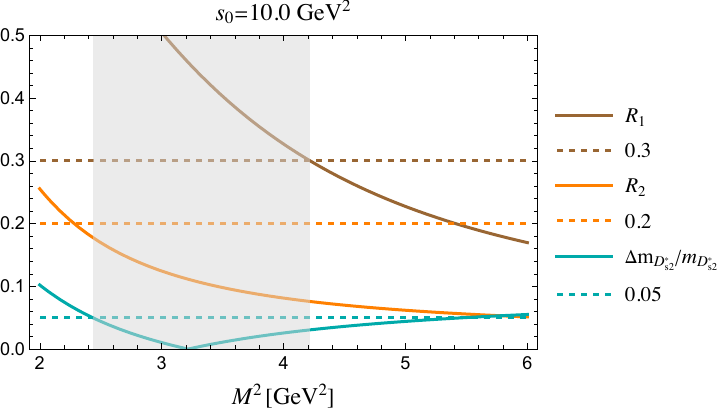}\\
\caption{\baselineskip 10pt  \small   Quantities  used to set the range of the Borel parameter $M^2$ in the two-point QCD sum rule in the case of  $D^*_2$ (left) and  $D^*_{s2}$ (right panels). The value of the continuum threshold $s_0$  is indicated above each panel. 
The continuous line $R_1$ is the ratio \eqref{ratioR}, the line $R_2$ is the ratio \eqref{ratioR1},  the line $\Delta m_{D^*_{(s)2}}/m_{D^*_{(s)2}}$ is drawn to enable the  criterion \eqref{crit3}, the dotted horizontal lines correspond to the values discussed in the text.  The shaded band indicates the range where all criteria are satisfied.
}\label{fig:criteriD}
\end{center}
\end{figure}

\begin{figure}[t]
\begin{center}
\includegraphics[width = 0.48\textwidth]{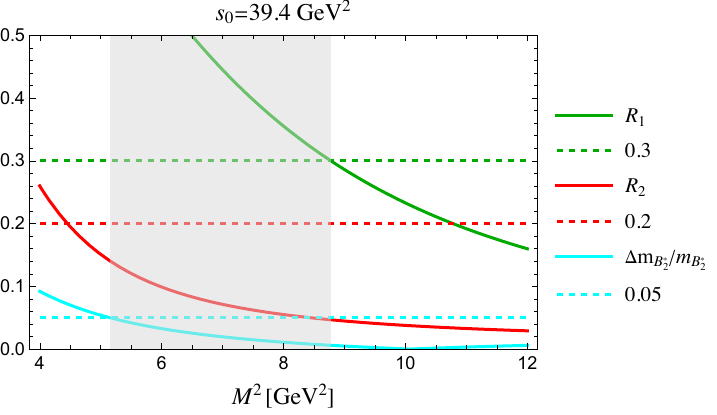} \hskip 0.2cm
\includegraphics[width = 0.48\textwidth]{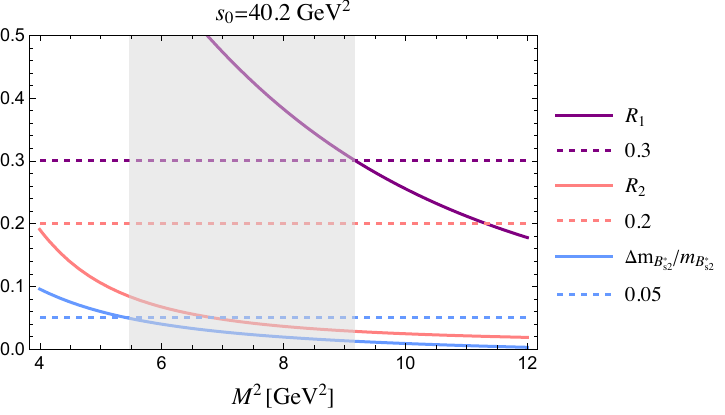}\\
\vskip 0.3cm
\includegraphics[width = 0.48\textwidth]{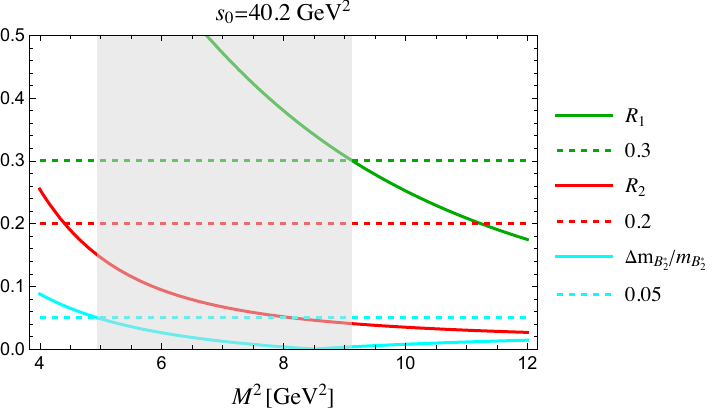} \hskip 0.2cm
\includegraphics[width = 0.48\textwidth]{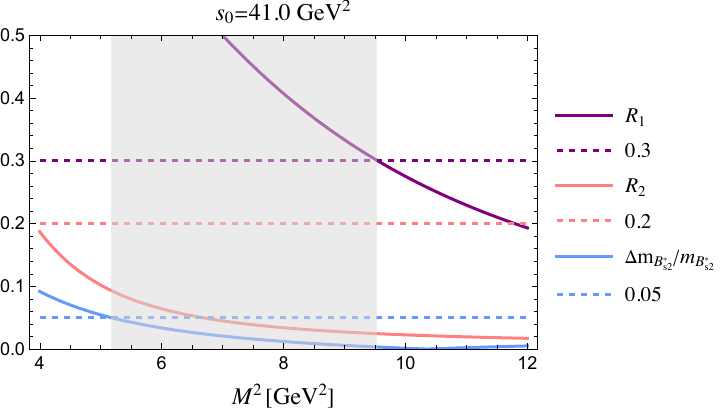}\\
\vskip 0.3cm
\includegraphics[width = 0.48\textwidth]{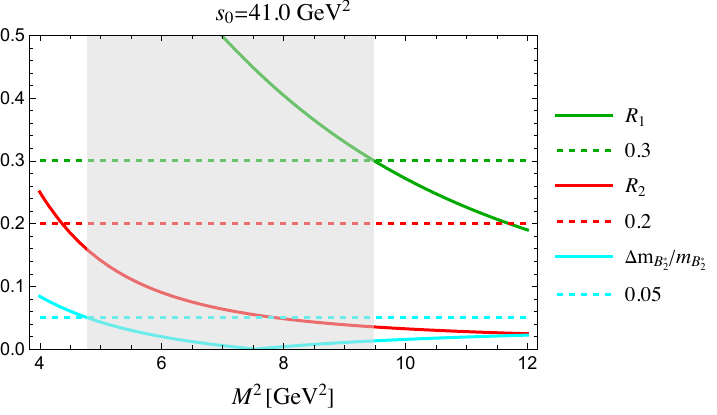} \hskip 0.2cm
\includegraphics[width = 0.48\textwidth]{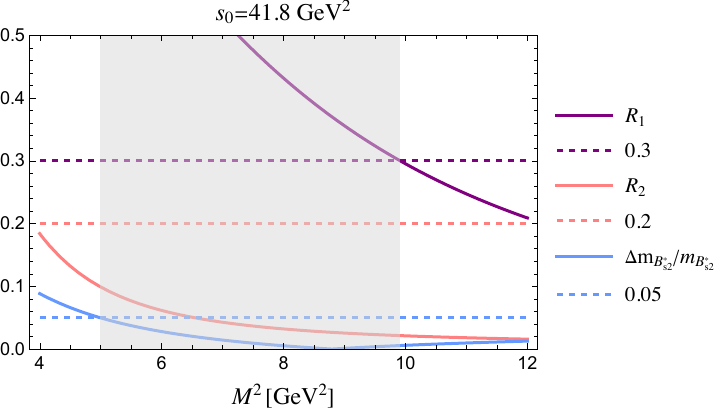}\\
\caption{ \baselineskip 10pt  \small   Quantities  used to set the range of the Borel parameter $M^2$ in the two-point QCD sum rule for  $B^*_2$ (left) and  $B^*_{s2}$ (right panels), with  notation  as in  
fig.~\ref{fig:criteriD}.}\label{fig:criteriB}
\end{center}
\end{figure}

In each case $H_2^* = D_2^*,\, D_{s2}^*,\, B_2^*,\, B_{s2}^*$ we determine
the ranges of $s_0$ and $M^2$ imposing  a set of conditions.
Considering the rhs of eq.~\eqref{sr} we look at the ratio of the result obtained when the upper limit of the dispersive  integral is $s_0$, and when it extends to larger values ($\simeq 3 s_0$)  approximating the full QCD contribution:
\be
R_1=\frac{\int_{(m_Q+m_q)^2}^{s_0} ds \, \rho^{\rm QCD}(s)\,e^{-s/M^2}+{\widehat \Pi}^{D=3}(M^2) +{\widehat \Pi}^{D=5}(M^2)}{\int_{(m_Q+m_q)^2}^{3s_0}ds \, \rho^{\rm QCD}(s)\,e^{-s/M^2}+{\widehat \Pi}^{D=3}(M^2) +{\widehat \Pi}^{D=5}(M^2)} \label{ratioR} \,\,.
\ee
To ensure  the dominance of the $H_2^*$ pole we require  $R_1  \gtrsim 0.3$, i.e. the lightest pole contribution amounting to at least $30\%$ of the full QCD result.
Moreover, we look at  the hierarchy  among the OPE terms in the sum rule. Specifically, for the ratio
\be
 R_2=\frac{{\widehat \Pi}^{D=5}(M^2)}{\int_{(m_Q+m_q)^2}^{s_0} ds \,  \rho^{\rm QCD}(s)\,e^{-s/M^2}+{\widehat \Pi}^{D=3}(M^2) +{\widehat \Pi}^{D=5}(M^2)}
\label{ratioR1}
\ee
we require $R_2  < 0.2$.
In the derived sum rule for $m_{H^*_2}$ we further require that the mass obtained from the sum rule reproduces the measured value $m_{H^*_2}^{exp}$   within $\simeq 5\%$: 
\be
|m_{H^*_2}-m_{H^*_2}^{exp}|<0.05 \, m_{H^*_2}^{exp} \label{crit3}\,\,.
\ee
For  the threshold,  we  initially  scan  the range $s_0\in[m_{H^*_2}^2,\,(m_{H^*_2}+0.5 \,\,{\rm GeV})^2]$, a typical mass difference between the ground state and the first radial excitation, then we look for the ranges of  $M^2$ where all  the imposed requirements are satisfied, checking the stability  against variations of $M^2$.  
The results in correspondence of  the various constraints are shown in fig.~\ref{fig:criteriD} and \ref{fig:criteriB}  for the charm and beauty mesons. 

\begin{figure}[t]
\begin{center}
\includegraphics[width = 0.45\textwidth]{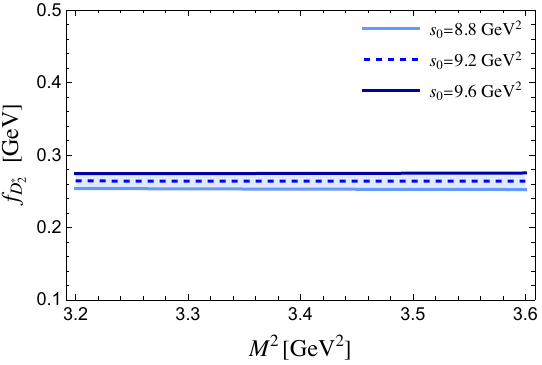} \hskip 0.3cm
\includegraphics[width = 0.45\textwidth]{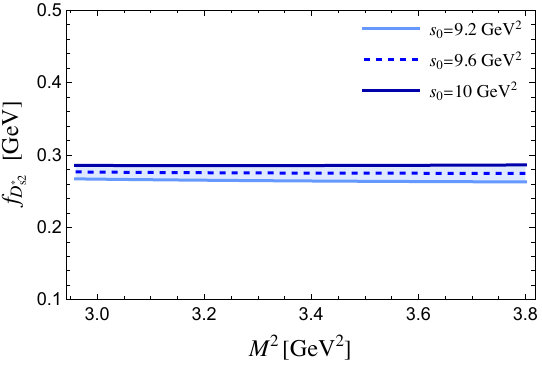} \\
\vskip 0.3cm
\includegraphics[width = 0.45\textwidth]{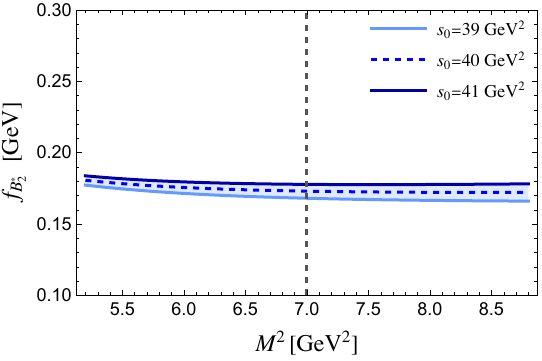} \hskip 0.3cm
\includegraphics[width = 0.45\textwidth]{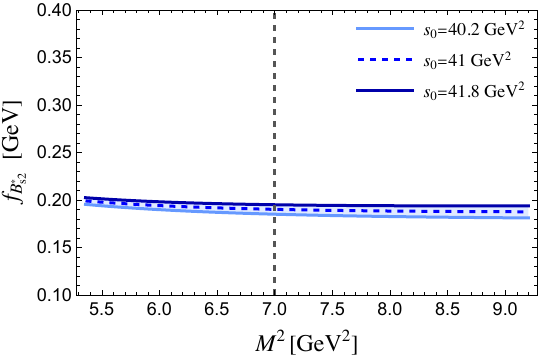}
\caption{\baselineskip 10pt  \small   Decay constant $f_{H_2^*}$ for $D_2^*$ and  $D_{s2}^*$ (left and right top panels),  and  $B_2^*$ and  $B_{s2}^*$ (left and right  bottom panels), varying the Borel parameter $M^2$. The curves correspond to the maximum, central and  minimum  value of the  continuum  threshold $s_0$  in table~\ref{tab:par}. For beauty mesons the vertical dotted line indicates the  lowest value chosen for $M^2$.
 The shaded bands show the uncertainties associated with the variations of the Borel parameter $M^2$ and the continuum threshold $s_0$. The final uncertainties quoted in table~\ref{tab:par} further include the effects of the renormalization-scale dependence of the heavy-quark masses and condensates.
}\label{fig:fT}
\end{center}
\end{figure}

The resulting ranges of parameters   are quoted in table~\ref{tab:par} together with the obtained decay constants.  The stability of $f_{H^*_2}$  against variation of the Borel parameter $M^2$ is shown in fig.~\ref{fig:fT}. 
For charmed mesons the stability  is found in the whole range
determined by the various criteria, for  beauty mesons  stability is achieved  further restricting  $M^2 \gtrsim 7$ GeV$^2$.

 The errors quoted in table~\ref{tab:par} originate from two different sources. The first one is obtained from the variation of the Borel parameter and continuum threshold within the working windows determined above. Since the perturbative spectral densities are evaluated at leading order in $\alpha_s$, we also estimate the sensitivity to the renormalization scale entering the running heavy-quark masses. To this end, we vary $m_c(\mu)$ in the range $1.20 - 1.34$ GeV and $m_b(\mu)$ in the range $4.0 - 4.4$ GeV, evolving the quark condensates consistently according to the renormalization-group relation $m_q(\mu) \langle {\bar q}q \rangle (\mu)=const$ . This produces a  variation in the result represented by  the second error quoted in table~\ref{tab:par}.

 The computed decay constants satisfy the scaling rule 
\be
\frac{f_{B^*_{(s)2}}}{f_{D^*_{(s)2}}} = \frac{\sqrt{m_{D^*_{(s)2}}}}{\sqrt{m_{B^*_{(s)2}}}} 
\ee
expected in  the heavy quark limit: using the central values of $f_{H^*_2}$ we obtain 
$f_{B^*_{2}}/f_{D^*_{2}}= 0.65$ and $  f_{B^*_{s2}}/f_{D^*_{s2}} = 0.686$, vs
$\sqrt{m_{D^*_{2}}}/\sqrt{m_{B^*_{2}}} = 0.655$ and 
$\sqrt{m_{D^*_{s2}}}/\sqrt{m_{B^*_{s2}}} = 0.663$.

Our result for $f_{D^*_2}$ is consistent with the one in  \cite{Colangelo:1992kc}.
As for other QCD sum rule analyses, a different interpolating current for $D^*_2$ is used in \cite{Sundu:2012zz}. In \cite{Wang:2014yza} two  interpolating current are considered, the current adopted here  and a current  with the covariant derivative replaced by the ordinary derivative. We agree on the perturbative term  after expanding our expression \eqref{pert2pt} to ${\cal O}(m_q)$,  we disagree on the nonperturbative contributions.

  We  comment  on the effect of the finite width of the tensor resonances on the results. In the derivation of the sum rule we have adopted the  narrow-width approximation, replacing the hadronic spectral function by a single-particle pole \cite{Uhlemann:2008pm}. This approximation is expected to be reliable for the members of the heavy-light $j_\ell^P=3/2^+$ doublet, which are relatively narrow states. Their  measured widths, reported by the PDG \cite{ParticleDataGroup:2024cfk}, are collected  in Table \ref{tab:par}. The ratios $\Gamma_{H_2^*}/m_{H_2^*}$ are approximately $1.9\times10^{-2}, \,6.6\times10^{-3},\, 3.5\times10^{-3}$, and $2.6\times10^{-4}$ for $D_2^*(2460), \,D_{s2}^*(2573),\, B_2^*$, and $B_{s2}^*$, respectively. 
A detailed treatment of the finite-width effects would require a dedicated analysis  beyond the scopes of this  work. The cases of the $\pi \pi$ and $K \pi$ light resonances have been treated in \cite{Cheng:2017smj,Descotes-Genon:2019bud}, but for them the ratio  $\Gamma/M$ is much larger. For the axial-vector and scalar charmed mesons belonging  to the $j_\ell^P=1/2^+$ spin  doublet,  finite-width corrections have been considered in \cite{Gubernari:2022hrq,Gubernari:2023rfu}.
On the other hand, the narrow width approximation has been adopted in the LCSR analysis of the semileptonic $B\to D_1$ transition in \cite{Gubernari:2022hrq}, where the $D_1$ meson belongs to the $j_\ell^P=3/2^+$ heavy-quark spin doublet.
To estimate the finite-width effects, we replaced the pole contribution in Eq. \eqref{srborel} by a Breit-Wigner distribution with constant width. The resulting correction can be expressed through the replacement $e^{-m_{H_2^*}^2/M^2}\to \varepsilon(M^2)$ \cite{Bruch:2004py,Gelhausen:2014jea,Cheng:2017smj,Descotes-Genon:2019bud}, where
\be
\varepsilon(M^2)=\frac{1}{\pi} \int_{s_{\rm th}}^\infty ds \, e^{-s/M^2} \frac{\sqrt{s} \, \Gamma(H^*_2)}{(s-m_{H_2^*}^2)^2+s \,\Gamma(H^*_2)^2}  \label{BW}
\ee
 is the integral of the  Breit-Wigner distribution and $s_{\rm th}$ is the lowest hadronic threshold (e.g. $m_{D^+} + m_{\pi^-}$, $m_{D^+}+m_{K^0}$, $m_{B^+} +m_{\pi^-}$ and $m_{B^+}+m_{K^0}$ in the four cases). In the Borel windows adopted in this work, the  decay constants are modified by the factor $\big(e^{-m_{H_2^*}^2/M^2}/\varepsilon(M^2)\Big)^{1/2}$, which  is about 1.014 for $D_2^*$, 1.009 for $D_{s2}^*$, 1.007 for $B_2^*$, and 1.002 for $B_{s2}^*$, with  a negligible dependence on $M^2$. Therefore, finite-width effects are well below the other theoretical uncertainties of the sum rule calculation.
We have also verified that the inclusion of the finite width has a negligible impact on the determination of the Borel windows and of the continuum thresholds, according to the criteria discussed in this section. Two of the criteria used to define the working regions do not concern the hadronic side of the sum rule,  the third criterion based on the ratio of the derivative sum rule to the original one is quite  unaffected  if  the pole term is replaced by the Breit-Wigner distribution.

\section{LCSR calculation of $B_q \to D_{q2}^*$ form factors}\label{sec:LCSR}
\subsection{Generalities}
To compute the $B_q \to D_{q2}^*$ form factors we  apply  LCSR  with $B$ LCDA. In the following we omit the indication of the light quark: $\bar B \equiv B_{q}$,  $H_2^* \equiv D_{q2}^*$. 

The starting point is  the correlator  between the vacuum and the $\bar B$ external state:
\be
{\cal F}_{\alpha \beta M}(p,q)=i \, \int d^4x e^{i p \cdot x}  \langle 0| T\{J_{\alpha \beta}(x) J_M(0) \} |{\bar B}(p+q) \rangle\,\,, 
\label{LCcorr}
\ee
with $p+q$ the $\bar B$ momentum.
$J_{\alpha \beta}$ is the  current  in \eqref{current} with $Q \equiv c$.
The  current $J_M={\bar c}\Gamma_M b$, with $M$ a set of Lorentz indices,   induces the  semileptonic $B \to H_2^*$  transition in the general low-energy  Hamiltonian governing 
$b \to c \ell \bar \nu_\ell$.
In  the SM Hamiltonian one has $\Gamma_M=\gamma_\mu$ and $\Gamma_M=\gamma_\mu \gamma_5$, in the general Hamiltonian one  also has $\Gamma_M=i\,\gamma_5$ and $\Gamma_M=\sigma_{\mu \nu}\gamma_5$. The scalar current does not contribute to  the  transition we are considering. 
The   hadronic matrix elements of the weak current  are parametrized in terms of  form factors:
\bea
\langle H_2^*(v^\prime,\epsilon) | \bar{c}  \gamma_\mu  b | \bar B(v) \rangle &=&\sqrt{m_{H_2^*} \, m_B} \,k_V(w)\,
\epsilon_{ \mu \alpha  \beta \sigma} {\tilde \epsilon}^{*\alpha} v^\beta v^{\prime \sigma}  \nn \\
\langle H_2^*( v^\prime,\epsilon) | \bar{c}  \gamma_\mu  \gamma_5 b | \bar B(v) \rangle &=&-i \,\sqrt{m_{H_2^*} \, m_B }\,
\Big[k_{A_1}(w) \,{\tilde \epsilon}_\mu^* +{\tilde \epsilon}_\alpha^*v^\alpha  \left(k_{A_2}(w) v_\mu +k_{A_3}(w)v^\prime_\mu \right) \Big]  \nn \\
\langle H_2^*(v^\prime,\epsilon) | \bar{c} \, i \gamma_5 b | \bar B (v) \rangle &=&\sqrt{m_{H_2^*} \, m_B} \, k_P(w){\tilde \epsilon}_\alpha^*v^\alpha  \nn \\
\langle H_2^*(v^\prime,\epsilon) | \bar{c}   \sigma_{\mu\nu} \gamma_5  b | \bar B(v) \rangle &=& \sqrt{m_{H_2^*} \, m_B} \, \Big[ k_{T_1}(w) ( {\tilde \epsilon}_\mu^*  v_\nu - {\tilde \epsilon}_\nu^*  v_\mu)  
+ k_{T_2}(w) ( {\tilde \epsilon}_\mu^*  v'_\nu - {\tilde \epsilon}_\nu^*  v'_\mu) \nn \\ 
&+& k_{T_3}( w) {\tilde \epsilon}_\alpha^* v^\alpha  ( v_\mu v'_\nu - v_\nu v'_\mu ) \Big] \,\,\, ,  \label{ff:chic2}
\eea
where $v$ and $v^\prime$ are 4-velocities, $p+q=m_B v$,  $w=v \cdot v^\prime$   and $\tilde \epsilon_\alpha=\epsilon_{\alpha \tau}v^\tau$. 

The correlator  \eqref{LCcorr} has a hadronic representation which involves  the form factors, 
and a QCD OPE representation expressed in terms of  $B$ meson LCDA. The form factors can be determined by matching the two representations. 

The hadronic representation is  a dispersion relation involving the imaginary part of the correlation function. In the single-particle + continuum model this is written as
\be
\frac{1}{\pi} {\rm Im}\, {\cal F}_{\alpha \beta M}(s)= \langle 0|J_{\alpha \beta}|H_2^* \rangle \langle H_2^*|J_M|{\bar B} (p+q) \rangle \delta(s-m_{H^*_2}^2)+ \rho^h_{\alpha \beta M}(s) \theta(s-s_0) \,\, , \label{LChad}
\ee
with $p^2=s$, and $\rho^h(s)$ comprising the  contributions above the hadronic threshold $s_0$. For the various $\Gamma_M$  the Lorentz structures allowing to access the  form factors can be identified:
\bea
{\cal F}_{\alpha \beta \mu}^{(V)}&=& {\cal F}_V \, \,\epsilon_{\tau \sigma \theta \mu} v^\tau v^{\prime \sigma} v^\delta (g^\theta_\beta g_{\delta \alpha} +g^\theta_\alpha g_{\delta \beta})+ {\rm other \,\,Lorentz\,\,  structures}\nn \\
{\cal F}_{\alpha \beta \mu}^{(A)}&=& {\cal F}_{A1} \, \,g_{\mu \alpha} v_\beta +{\cal F}_{A2} \, \,v_\mu v_\alpha v_\beta+{\cal F}_{A3} \, \,v_\mu^\prime v_\alpha v_\beta
+ {\rm other \,\,Lorentz\,\,  structures} \nn \\
{\cal F}_{\alpha \beta }^{(P)}&=&{\cal F}_{P} \, \, v_\alpha v_\beta+ {\rm other \,\,Lorentz\,\,  structures} \nn \\
{\cal F}_{\alpha \beta \mu \nu }^{(T)}&=& {\cal F}_{T1} \, \, v_\alpha (g_{\mu \beta} v_\nu -g_{\nu \beta} v_\mu)
 +{\cal F}_{T2} \, \,v_\alpha (g_{\mu \beta} v_\nu^\prime -g_{\nu \beta} v_\mu^\prime)
 +{\cal F}_{T3} \, \, v_\alpha v_\beta (v_\mu v_\nu^\prime -v_\nu v_\mu^\prime ) \nn \\
&+&{\rm other \,\,Lorentz\,\,  structures}.  \label{structures}
\eea
In each case the contribution of the first term in the rhs of eq.~\eqref{LChad} is written as ${\cal F}_F=c_F^{had} k_F$, where $F \equiv V,\,A_{1,2,3},\,P,\,T_{1,2,3}$. The coefficient $c_F^{had}$, which depends on the considered form factor,  is  given  in appendix \ref{appB}.
The hadronic representation  reads: 
\be
{\cal F}_{F}(p^2,q^2)=\frac{c_F^{had} \, k_F(q^2)}{m_{H_2^*}^2-p^2}+\int_{s_0}^\infty ds \,  \frac{\rho^h_{F}(s,q^2)}{s-p^2}\,\,. \label{hadF}
\ee
To obtain the  QCD representation of the correlator,  the $c$ quark propagator  $S_c(x,0)$  in  eq.~\eqref{LCcorr} is reconstructed:
\be
{\cal F}_{\alpha \beta M}(p,q)=i \, \int d^4x e^{i p \cdot x}  \langle 0| {\bar q}(x)\Big[i A_{\alpha \beta}^{\xi \psi}\DD_\xi \, \gamma_\psi \Big] \,i \, S_c(x,0) \Gamma_M b(0)|{\bar B}(p+q) \rangle \,\, . \label{OPE1}
\ee
The propagator  is expanded up to one-gluon insertion:
\be
i\, S_c(x,0)=i \, S_c^{(0)}(x,0)+i \,S_c^{(1)}(x,0) \,\, ,
\ee
with
\bea
i \, S_c^{(0)}(x,0)&=& i \int \frac{d^4 p_1}{(2 \pi)^4}e^{-ip_1 \cdot x} \frac{{\spur p}_1+m_c}{p_1^2-m_c^2} \\
i \, S_c^{(1)}(x,0)&=& i \int \frac{d^4 p_1}{(2 \pi)^4}e^{-ip_1 \cdot x}\int_0^1 du \,G_{\tau \sigma}(ux) \Big[\frac{u\, x^\tau \gamma^\sigma}{p_1^2-m_c^2}-\frac{({\spur p}_1+m_c)\sigma^{\tau \sigma}}{2(p_1^2-m_c^2)^2}\Big] \,\,.
\eea
The  $B$ to  vacuum matrix element  can be written in terms of  increasing twist LCDA  of the $B$ meson. For this purpose  the $b$ quark field is replaced by  the heavy field $h_v$ defined in the heavy quark effective theory.
Two-particle matrix elements
\be
\langle 0|{\bar q}(x) h_v(0)|{\bar B}(p+q)\rangle_{\alpha_1 \alpha_2} \label{2pt0}
\ee
and three-particle matrix elements 
\be
\langle 0|{\bar q}(x) g_s G_{\tau \nu} (u\,x) h_v(0)|{\bar B}(p+q) \rangle_{\alpha_1 \alpha_2} \,\,\,  \label{3pt0}
\ee
are therefore involved,  with $\alpha_1,\alpha_2$   Dirac indices.
Such matrix elements  can be expressed in terms of  LCDA of the $B$ meson. The two-particle matrix element reads
\bea
\langle 0|{\bar q}(x) h_v(0)|{\bar B}(p+q)\rangle &=&-i \frac{f_B m_B}{4} \int_0^\infty d\omega \,\Big\{ (1+{\spur v})\Big[\phi_+(\omega)-g_+(\omega) \partial_\lambda \partial^\lambda \nn \\
&&+\frac{1}{2}\Big({\bar \phi}_\pm (\omega)-{\bar G}_\pm(\omega)\partial_\lambda \partial^\lambda \Big) {\spur \partial} \Big]\gamma_5 \Big\}\, e^{-i \ell \cdot x} \Big|_{\ell=\omega \,v} \,\,\, , \qquad \label{2pt1}
\eea
with $f_B$ the $B$ meson decay constant, $\partial_\lambda=\displaystyle\frac{\partial}{\partial \ell^\lambda}$\,\,   and   \cite{Gubernari:2022hrq}
\be
{\bar \phi}_\pm (\omega)=\int_0^\omega d\tau \, [\phi_+(\tau)-\phi_-(\tau)] \,\,\, , \hskip 1 cm  {\bar G}_\pm (\omega)=\int_0^\omega d\tau \, [g_+(\tau)-g_-(\tau)] \,\,. \label{barred}
\ee
The three-particle matrix element is written as 
\bea
\langle 0| \bar{q}(x) g_s G_{\tau\sigma}(u \, x) h_{v}(0) |\bar{B}(p+q) \rangle &=&
 \frac{f_B m_B}{4} \int_0^\infty d\omega_1 \int_0^\infty\ d\omega_2 \, e^{-i(\omega_1 + u \omega_2)\, v\cdot x}  \nn \\
&&\hskip -5cm \bigg\{(1+\spur{v})\bigg[(v_\tau\gamma_\sigma-v_\sigma\gamma_\tau) \big[\psi_A-\psi_V \big] 
        -i\sigma_{\tau\sigma }\psi_V  \nn \\
    &&\hskip -5cm + (\partial_\tau v_\sigma-\partial_\sigma v_\tau) \bar{X}_A
      - (\partial_\tau\gamma_\sigma-\partial_\sigma\gamma_\tau) [\bar{W} + \bar{Y}_A]
      + i\epsilon_{\tau\sigma\alpha\beta}\partial^\alpha v^\beta\gamma_5 \bar{\tilde{X}}_A \nn\\
    &&\hskip -5cm  - i\epsilon_{\tau\sigma\alpha\beta}\partial^\alpha \gamma^\beta\gamma_5 \bar{\tilde{Y}}_A
      - u (\partial_\tau v_\sigma-\partial_\sigma v_\tau){\spur \partial} \bar{\bar{W}}
      + u (\partial_\tau \gamma_\sigma-\partial_\sigma \gamma_\tau){\spur \partial} \bar{\bar{Z}}
    \bigg]\gamma_5\bigg\}\, , \qquad \label{3pt1}
\eea
with $\partial_\lambda=\displaystyle\frac{\partial}{\partial \ell_1^\lambda}$ and $\ell_1=(\omega_1 + u\, \omega_2)\, v$.
In eq.~\eqref{3pt1} we have omitted  the argument $(\omega_1,\,\omega_2)$ of the various LCDA, and   for a generic function $\psi(\eta_1,\eta_2)$ we have defined
\bea
{\bar \psi}(\omega_1,\,\omega_2)&=&\int_0^{\omega_1} d\eta_1 \, \psi (\eta_1,\,\omega_2) \nn \\
{\bar {\bar \psi}}(\omega_1,\,\omega_2)&=&\int_0^{\omega_1} d\eta_1 \int_0^{\omega_2} d\eta_2 \, \psi (\eta_1,\,\eta_2) \,\,.
\eea
The expressions of the $B$ meson   LCDA  in \eqref{2pt1}, \eqref{barred} and \eqref{3pt1}   are  in appendix \ref{appC}.

After computing  traces,  derivatives and   integrations, each LCDA appears combined with 
$1/[(p-\ell)^2-m_c^2]^k$. In the  two-particle contribution, since  $\ell=\omega \,v$ and   $\omega=m_B \sigma$, such terms are recast in the form
\be
 \frac{1}{[(p-\ell)^2-m_c^2]^k}=\frac{1}{{\bar \sigma}^k[p^2-s(\sigma,q^2)]^k} \,\,\, , 
 \ee
 where $\bar \sigma=1-\sigma$ and 
\be
s(\sigma,q^2)=\sigma m_B^2-\frac{\sigma q^2-m_c^2}{\bar \sigma}\,\,.
\ee
The same structure is recovered in the three-particle case,  since $\ell_1=(\omega_1+u \,\omega_2) v$ and $\ell_1=m_B \sigma v$.
 For each form factor $F$  the OPE side of the sum rule has the expression 
\be
{\cal F}^{OPE}_F(p^2,q^2)=c^{OPE}_F \sum_{i=1}^4 \int_0^{\infty}d \sigma 
\frac{{\bar {\cal F}}^{OPE(F)}_i(\sigma)}{[p^2-s(\sigma,q^2)]^i} \,\,\, ,
\ee
with the various contributions grouped according to the power of the denominator. The  functions ${\bar {\cal F}}^{OPE(F)}_i(\sigma)$ contain both the contribution of  two- and three-particle LCDA  upon integration of  the latter ones over $\omega_1,\,\omega_2$:
\be
{\bar {\cal F}}^{OPE(F)}_i(\sigma)=\sum_{\psi^{\rm 2pt}} C_{\psi^{\rm 2pt}}^{i (F)} \, \psi^{\rm 2pt}(m_B \sigma)+\sum_{\psi^{\rm 3pt}} \int_0^{m_B \sigma} d\omega_1 \,  \int_{m_B \sigma -\omega_1}^\infty \frac{d\omega_2}{\omega_2}\, 
C_{\psi^{\rm 3pt}}^{i(F)} \, \psi^{\rm 3pt}(\omega_1,\omega_2)\,\,\,\, . \label{FOPE}
\ee

The subtraction of the continuum, obtained invoking quark-hadron duality,  can be done following  the procedure  outlined in \cite{Gubernari:2018wyi}.  The integral in \eqref{FOPE} is split in two regions separated by $\sigma_0=\sigma(s_0,q^2)$,  with $s_0$  the hadronic threshold above  $m_{H^*_2}$. 
For a generic function $f(\sigma)$ one has to compute integrals of the type 
\be
I_k=\int_0^\infty d \sigma \frac{f(\sigma)}{[p^2-s(\sigma,q^2)]^k}=I_{k,1}+I_{k,2} \,\, ,  \label{Ik}
\ee
with
\bea
I_{k,1}&=& \int_0^{\sigma_0}  d \sigma \frac{f(\sigma)}{[p^2-s(\sigma,q^2)]^k} \nn \\
I_{k,2}&=& \int_{\sigma_0} ^\infty d \sigma \frac{f(\sigma)}{[p^2-s(\sigma,q^2)]^k} \,\,\, . \label{Ik1}
\eea
For $I_{k,1}$ the Borel transformation with respect to $p^2$ gives
\be
\hat I_{k,1}=\int_0^{\sigma_0}  d \sigma f(\sigma) \frac{(-1)^k}{(k-1)!} \frac{e^{-s(\sigma,q^2)/M^2}}{(M^2)^{k-1}} 
\ee
with $M^2$   the Borel parameter.
$I_{k,2}$  can be rearranged  writing 
$
\displaystyle\frac{1}{[p^2-s(\sigma,q^2)]^k}$ in terms of the $(k-1)$th derivative of $\displaystyle \frac{1}{[p^2-s(\sigma,q^2)]}$, and  performing subsequent integrations by parts.
After Borel transformation one has
\bea
\hat I_{k,2}&=&  \frac{(-1)^k}{(k-1)!}\nn \\
&\times&\Bigg[ e^{-\frac{s(\sigma,q^2)}{M^2}}\sum_{j=1}^{k-1}\frac{1}{(M^2)^{k-j-1}}\frac{1}{s^\prime}\Big(
\frac{d}{d \sigma}\frac{1}{s^\prime} \Big)^{j-1}f(\sigma) \Big|_{\sigma=\sigma_0} 
+\int_{\sigma_0}^\infty d \sigma \, e^{-\frac{s(\sigma,q^2)}{M^2}}\Big(\frac{d}{d \sigma}\frac{1}{s^\prime} \Big)^{k-1}f(\sigma)\Bigg ] \nn \\ \label{Ik2}
\eea
where $s^\prime=\displaystyle\frac{d s(\sigma,\,q^2)}{d\sigma}$. The first term in  \eqref{Ik2} exists only for $k>1$. 
Invoking quark-hadron duality, the last term is identified with the  integral of the hadronic spectral function $\rho^h_F$ in eq.~\eqref{hadF}. 

\subsection{Structure of the LC sum rule}
After continuum subtraction and Borel transformation, the hadronic side of the sum rule reads:
\be
{\cal F}^{had}_F=c^{had}_F k_F \,e^{-\frac{m_{H^*_2}^2}{M^2}}\,\,
\ee
for the form factor $F=V,\,A_{1,2,3},\,P,\,T_{1,2,3}$.
Matching this with the OPE representation in eq.~\eqref{FOPE},  the LC sum rule  is obtained in the form
\vskip 1cm
\bea
k_F(q^2) &=&c_F \, e^{\frac{m_{H^*_2}^2}{M^2}}
\Bigg\{ \int_0^{\sigma_0}d\sigma e^{-\frac{s(\sigma,q^2)}{M^2}}\,\Bigg[-{\bar{ \cal F}}^{OPE(F)}_1+\frac{{\bar {\cal F}}^{OPE(F)}_2}{M^2}-\frac{{\bar{ \cal F}}^{OPE(F)}_3}{2(M^2)^2}+\frac{{\bar {\cal F}}^{OPE(F)}_4}{6(M^2)^3}\Bigg]\nn \\ 
&&+\frac{e^{-\frac{s(\sigma,q^2)}{M^2}}}{s^\prime(\sigma,q^2)}\Bigg[{\bar {\cal F}}^{OPE(F)}_2-\frac{1}{2}\Big(\frac{{\bar{ \cal F}}^{OPE(F)}_3}{M^2}+{\bar{ \cal F}}^{OPE(F)}_{3,1}\Big) \nn \\
&&+\frac{1}{6}\Big(\frac{{\bar{ \cal F}}^{OPE(F)}_4}{(M^2)^2}+\frac{{\bar{ \cal F}}^{OPE(F)}_{4,1}}{M^2}+{\bar{ \cal F}}^{OPE(F)}_{4,2}\Big)\Bigg]\Bigg|_{\sigma=\sigma_0}\Bigg\} \, , \label{sr-gen} 
\eea
where $c_F=c^{OPE}_F/c^{had}_F$ and
\bea
{\bar{ \cal F}}^{OPE(F)}_{i,1}&=&\frac{d}{d \sigma}\Big( \frac{1}{s^\prime}{\bar{ \cal F}}^{OPE(F)}_i \Big) \nn \\
{\bar{ \cal F}}^{OPE(F)}_{i,2}&=&\frac{d}{d \sigma}\Big( \frac{1}{s^\prime}{\bar{ \cal F}}^{OPE(F)}_{i,1} \Big) \,\,. \label{sr-gen1}
\eea
Separating the two- and  three-particle contributions  we write
\be
{\bar{ \cal F}}^{OPE(F)}_i(\sigma)={\bar{ \cal F}}^{2pt(F)}_i(\sigma)+{\bar{ \cal F}}^{3pt(F)}_i(\sigma) \,\, .\label{sr-gen2}
\ee
We also write
\be
{\bar{ \cal F}}^{3pt(F)}_i(\sigma)=\int_0^{m_B \sigma}\,d\omega_1\,\int_{m_B \sigma -\omega_1}^\infty \, 
\frac{d\omega_2}{\omega_2} \Big[{\bar {\cal P}}^{3pt(F)}_i(\sigma,\omega_1,\omega_2)+{\bar {\cal R}}^{3pt(F)}_i(\sigma,\omega_1,\omega_2)\Big]\,\,, \label{sr-gen3}
\ee
 splitting the three-particle contributions into ${\bar {\cal P}}^{3pt(F)}_i$ and ${\bar {\cal R}}^{3pt(F)}_i$, the latter one corresponding to  the gluon emitted from a vertex due to  the covariant derivative in the current  \eqref{current}.
The  expressions of  ${\bar{ \cal F}}^{2pt(F)}_i$,  ${\bar {\cal P}}^{3pt(F)}_i$ and ${\bar {\cal R}}^{3pt(F)}_i$ are in appendix \ref{appB}.

\subsection{$B_q \to D_{q2}^*$ form factors: results}\label{sec:Numerics}
 To  compute the  $B_q \to D_{q2}^*$ form factors we use the $B_q$ LCDA available in the literature together with a set of  numerical parameters, see the appendix \ref{appC}.
 The truncation scheme adopted in this work corresponds to keeping the contributions  consistent with twist-four accuracy in the light-cone power expansion of the correlation function. This implies that LCDAs of formally higher twist are included whenever they are required by Wandzura–Wilczek relations or equations-of-motion identities. They do not constitute independent higher-order non-perturbative inputs.
The continuum threshold $s_0$ is the same determined in the
two-point sum rule for charmed mesons. The Borel parameter is allowed to vary in the range $M^2\in [3,\, 4]$ GeV$^2$ for $B \to D_2^*$, and $M^2\in [2.8,\, 4]$ GeV$^2$ for $B_s \to D_{s2}^*$ form factors. 
For each form factor,  stability against variations of $M^2$ is  checked.
We evaluate each form factor using
eq.~\eqref{sr-gen} at large negative  $q^2$, where the method is more precise, determining the coefficients $a_i$ of the  parametrization
\be
k_F^{\rm fit}(q^2)=\frac{1}{1-\frac{q^2}{m_R^2}} \sum_{i=0}^{1} a_i \left[z(q^2)-z(0)\right]^i \label{kFfit}
\ee
based on the conformal mapping
\be
z(q^2)=\frac{\sqrt{t_+-q^2}-\sqrt{t_+-t_0}} {\sqrt{t_+-q^2}+\sqrt{t_+-t_0}}\,.
\ee
$t_+$ is $t_+=(m_{B_{(s)}}+m_{D_{(s)2}^*})^2$, and  
$t_0=(m_{B_{(s)}}+m_{D_{(s)2}^*})(\sqrt{m_{B_{(s)}}}-\sqrt{m_{D_{(s)2}^*}})^2$ is chosen to  ensure that $|z|$ is small in the kinematical region. Due to the quite narrow range of $q^2$,
the truncation of the $z$ expansion to $i=1$ appears sufficient, since we have
\be
\frac{\Big|a_1 \big[z(q^2)-z(0) \big] \Big|}{|a_0|} < 1 .
\ee
\begin{figure}[t]
\begin{center}
\includegraphics[width = \textwidth]{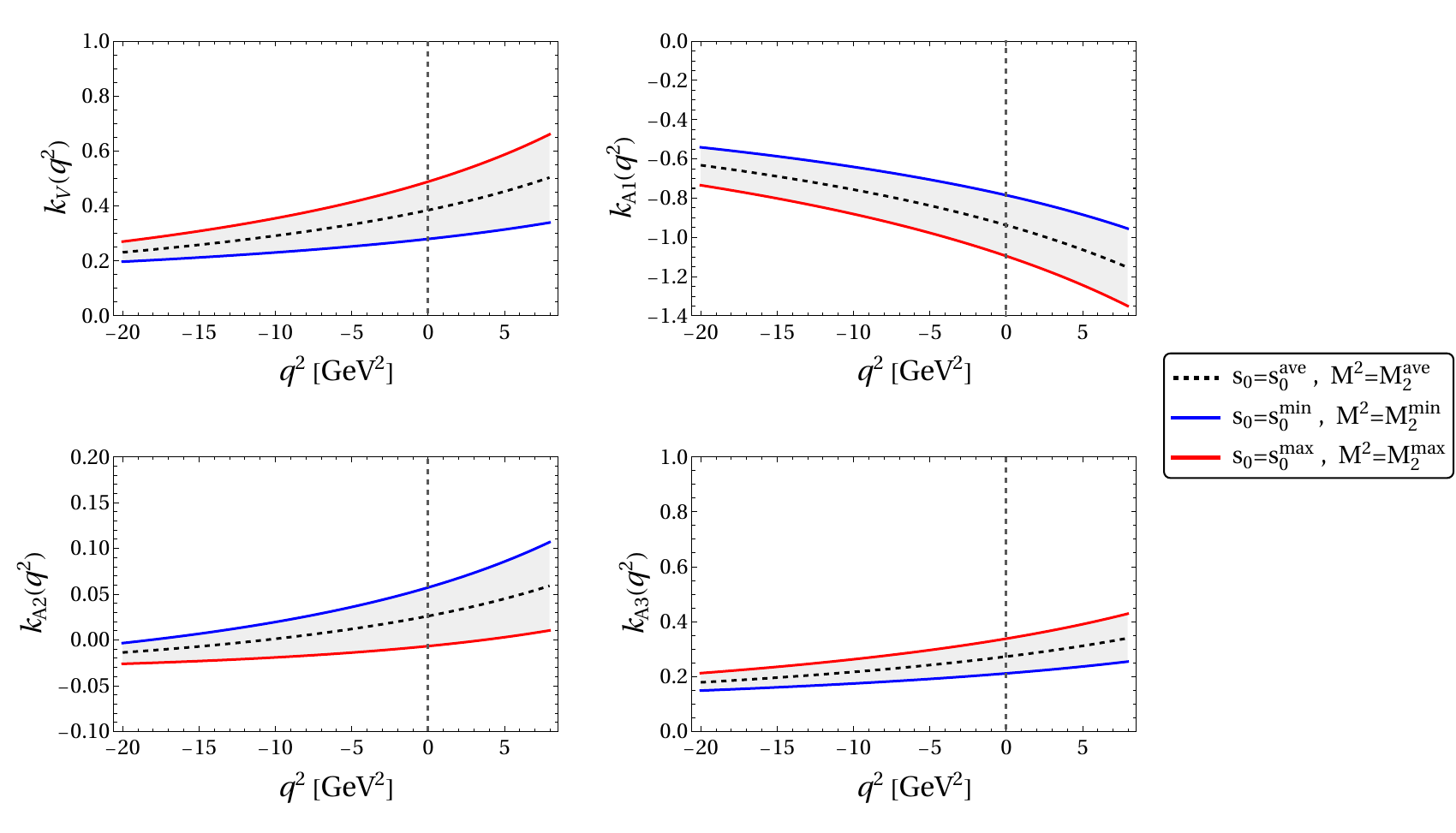}
\caption{\baselineskip 10pt  \small   $B \to D_2^*$  form factors of vector  and axial-vector current.
}\label{fig:FFSM}
\end{center}
\end{figure}
The  parameter $m_R$ can be identified with the mass of a $b\bar c$ state with  suitable quantum numbers: the  chosen value in the various cases  is in table~\ref{tab:a0a1}. The parameters $a_0$ and $a_1$ are determined
by a fit of form factor obtained from the sum rule.  The quoted uncertainties of the fit parameters are obtained by fitting the upper and lower boundaries of the LCSR uncertainty bands. The fitting procedure is adaptive,  
it depends on the behavior of the sum rule for the specific form factor.
We compute from eq.~\eqref{sr-gen} the values $k_F(q_k^2)$ for
$q_k^2\in[q^2_L,\,q^2_H]$  in steps of $0.5$ GeV$^2$. The uncertainty
$\delta k_F(q^2)\simeq (5\text{--}15)\times 10^{-2} \,k_F(q^2)$ is assigned to the
computed form factors, estimated from the variation of the results changing the threshold and the Borel parameter. 
The $\chi^2$ function
\be
\chi_F^2=\sum_k \frac{\left[k_F(q_k^2)-k_F^{\rm fit}(q_k^2)\right]^2} {\left[\delta k_F(q_k^2)\right]^2}
\ee
is minimized to determine the parameters $a_0$ and $a_1$.
The fitting  $q^2$ range is determined through an adaptive procedure: The lower value $q_L^2$ is chosen as either
$-25$ or $-20$ GeV$^2$, while the upper value  $q_H^2$ is varied in the range $[-14,-4]$ GeV$^2$. 
The range is chosen to avoid the region close to the hadronic threshold and to ensure the reliability of the light–cone expansion. For each choice of the range
$[q_L^2,q_H^2]$ we minimize the $\chi_F^2$ function and select the range providing the best fit quality. We have verified that the
resulting fit parameters remain stable under  variations of the fitting interval, as discussed below.
%
\begin{table}[b!]
\centering
\begin{tabular}{cccccc}
\hline \hline
 form factor & $m_R  \hskip 0.1cm [GeV]$ &$a_0$ & $a_1$ & $a_0^s$ & $a_1^s$  \\
\hline
$k_V$     &$6.33$& $0.385 \pm 0.105$  &$-0.530 \pm^{0.630}_{0.749}$&$0.646 \pm_{0.127}^{0.124} $&$-1.721\pm^{0.995}_{0.751}$\\
$k_{A_1}$&$6.74$&\hskip -0.3cm $-0.938\pm^{0.158}_{0.153}$ &\hskip 0.3cm $0.381 \pm^{0.141}_{0.297}$&\hskip -0.2cm $-1.359 \pm^{0.155}_{0.166}$ &\hskip 0.3cm $0.775\pm^{0.262}_0$\\
$k_{A_2}$&$6.74$& $0.026\pm^{0.031}_{0.033}$  &$-0.638\pm_{0.229}^{0.208}$&$ 
0.026 \pm^{0.057}_{0.052}$ &$ -1.061\pm_{0.526}^{0.392}$\\
$k_{A_3}$&$6.74$&$0.273 \pm 0.065$ &$ -0.210\pm^{0.254}_{0.234}$&
$0.460 \pm^{0.081}_{0.076}$&$-0.808\pm^{0.361}_{0.304}$\\
$k_P$      &$6.274$&$0.561 \pm_{0.093}^{0.097} $ &$-2.160\pm^{0.533}_{0.485}$&
$ 0.797 \pm^{0.103}_{0.095}$&$ -3.370\pm^{0.576}_{0.492}$ \\
$k_{T_1}$&$6.33$&\hskip -0.2cm $0.408\pm^{0.039}_{0.034}$ &$-2.476 \pm^{0.280}_{0.312}$&
$ 0.510\pm^{0.082}_{0.060}$ &$ -2.912\pm^{0.624}_{0.474}$\\
$k_{T_2}$&$6.33$&$0.385 \pm 0.105$ &$-0.530 \pm^{0.630}_{0.749}$ &$ 0.646 \pm^{0.124}_{0.127}$ &$-1.721\pm^{0.995}_{0.751}$\\
$k_{T_3}$&$6.33$&\hskip -0.3cm $-0.159 \pm 0.030$ &\hskip 0.3cm $1.219 \pm 0.250$&\hskip -0.2cm 
$ -0.226 \pm 0.033$&\hskip 0.3cm $ 1.795\pm 0.284$\\
\hline
\hline
\end{tabular}
\caption{\baselineskip 10pt  \small  Parameters in  \eqref{kFfit}.   $a_{0,1}^{(s)}$  correspond to  each  form factor of $B \to D^*_2$  (first two columns) and  $B_s \to D^*_{s2}$ (last two columns).}
\label{tab:a0a1}
\end{table}

\begin{figure}[t]
\begin{center}
\includegraphics[width = \textwidth]{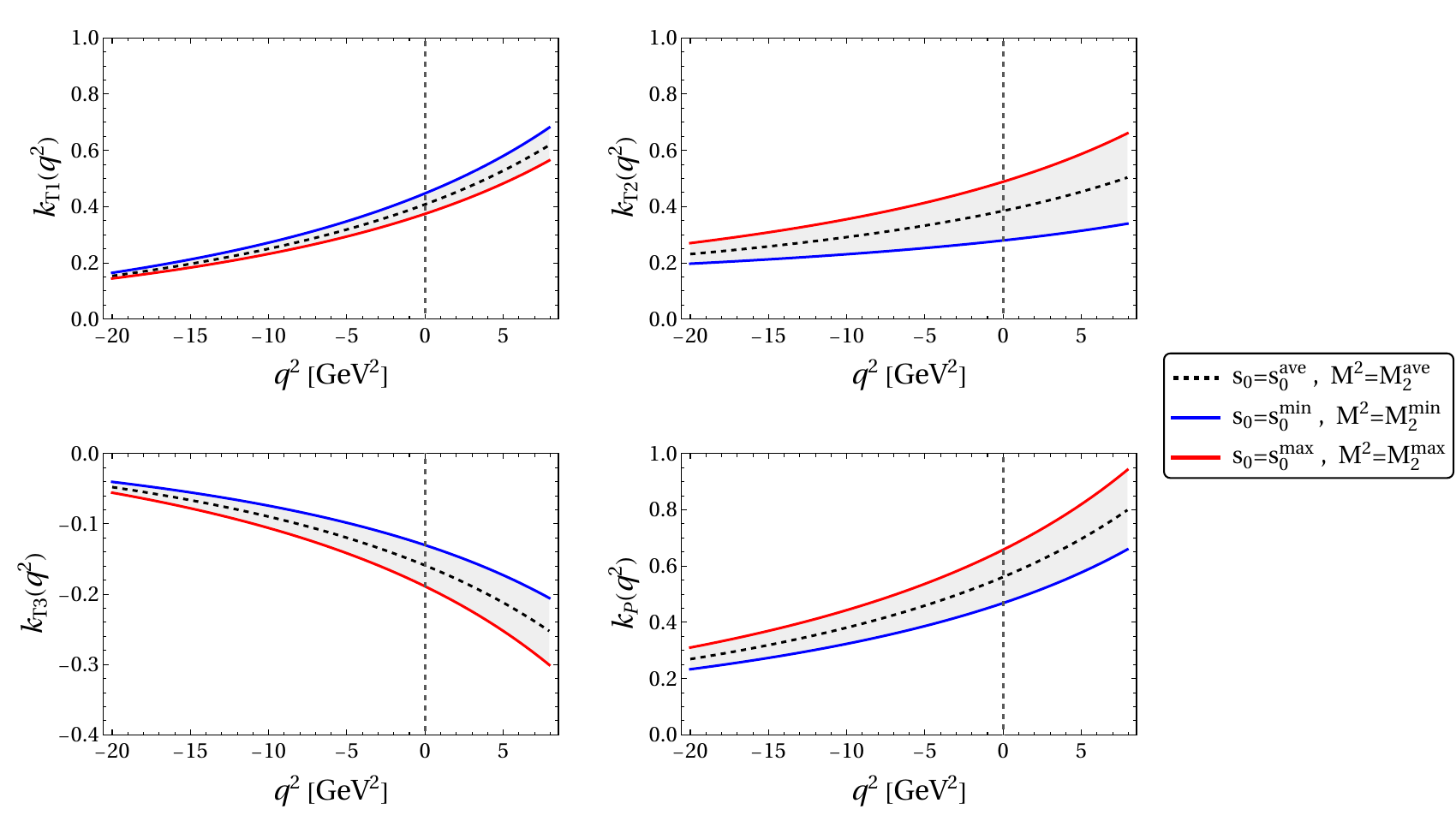}
\caption{\baselineskip 10pt  \small    $B \to D_2^*$  form factors of tensor and pseudoscalar  current.
}\label{fig:FFBSM}
\end{center}
\end{figure}

\begin{figure}[t]
\begin{center}
\includegraphics[width = \textwidth]{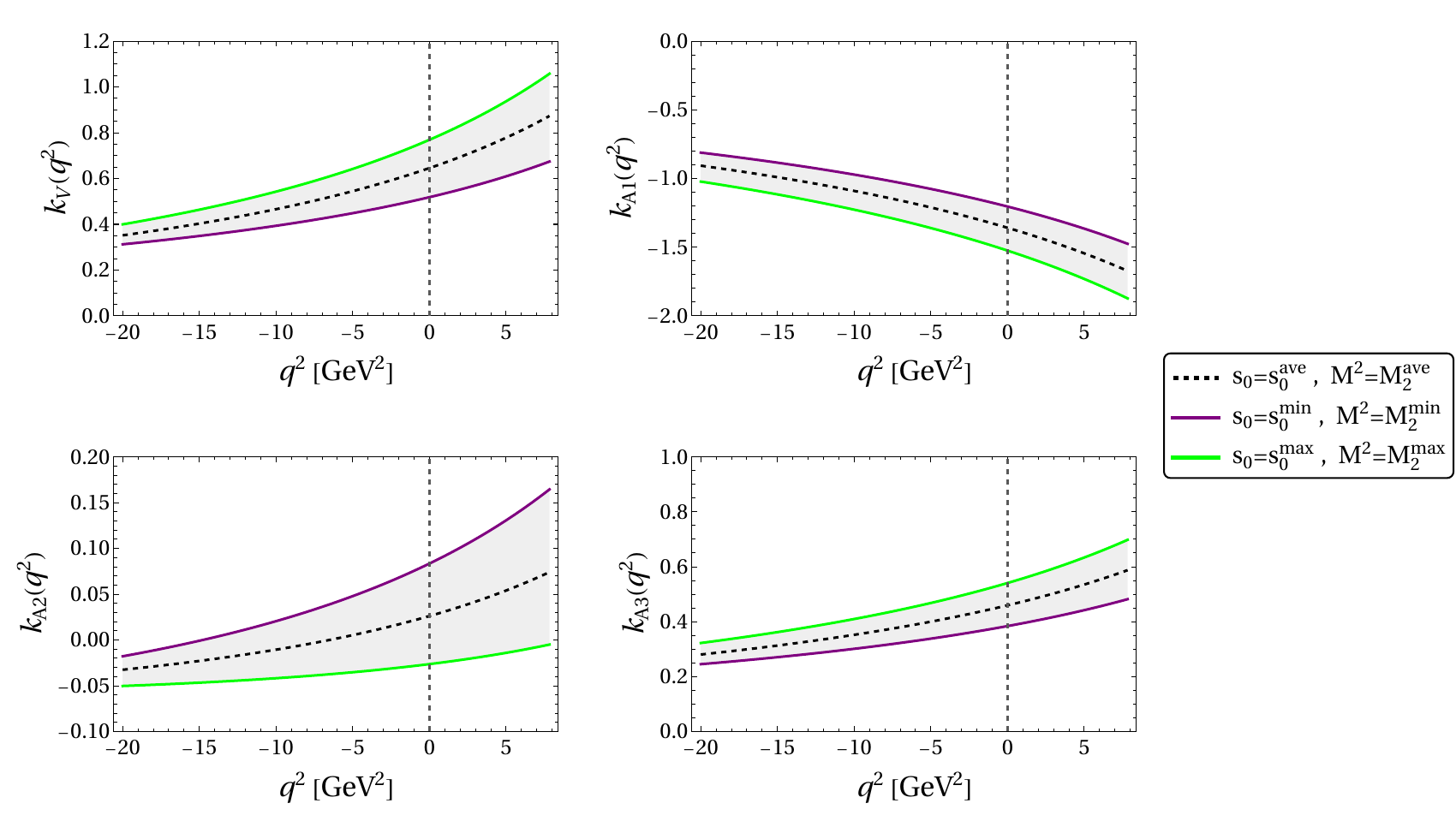}
\caption{\baselineskip 10pt  \small    $B_s \to D_{s2}^*$ form factors of vector  and axial-vector current.
}\label{fig:FFSMBs}
\end{center}
\end{figure}

\begin{figure}[b]
\begin{center}
\includegraphics[width = \textwidth]{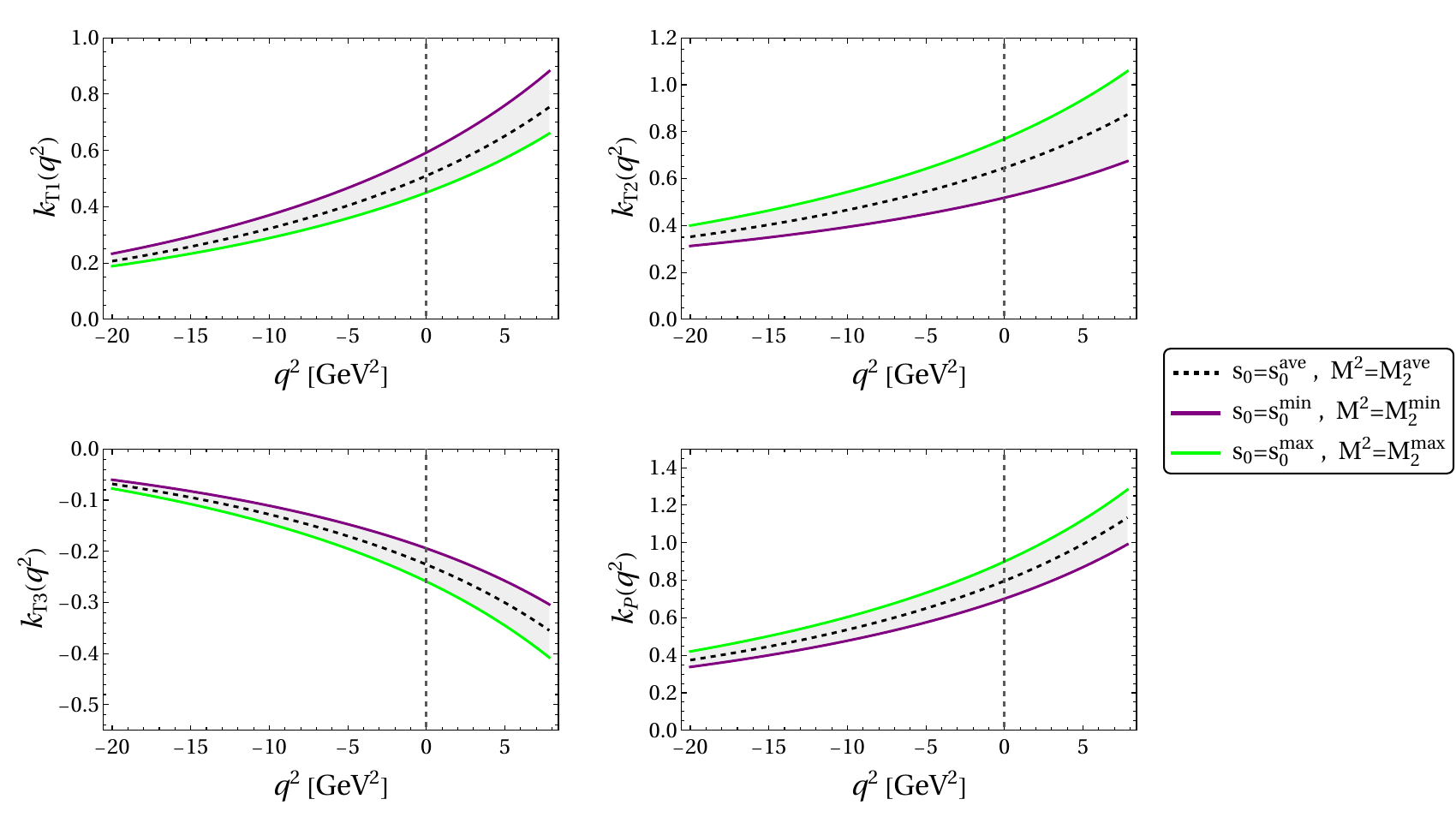}
\caption{\baselineskip 10pt  \small     $B_s \to D_{s2}^*$  form factors of tensor and pseudoscalar  current.
}\label{fig:FFBSMBs}
\end{center}
\end{figure}

In figs.~\ref{fig:FFSM}, \ref{fig:FFBSM} and  \ref{fig:FFSMBs}, \ref{fig:FFBSMBs} we display the computed  $B \to D^*_2$ and  $B_s \to D^*_{s2}$ form factors.
 The uncertainties of the fitted form factors are obtained by propagating the uncertainties of the LCSR predictions. Besides the variations of the Borel parameter and continuum threshold, we include the uncertainty associated with the renormalization scale dependence of the heavy quark masses through the corresponding decay constants. For the  $B$ meson LCDA, the reference scale is taken to be $\mu =1$ GeV, and the corresponding uncertainty is estimated by evolving the leading-twist parameter $\lambda_B$ to $\mu =1.5$ GeV using its renormalization-group evolution \cite{Braun:2003wx}. The effects induced by the higher-twist parameters $\lambda_E^2$ and $\lambda_H^2$ are found to be numerically negligible and are therefore not included in the final error budget. The resulting uncertainties are propagated to both fit parameters $a_0$ and $a_1$, whose values are reported in  table~\ref{tab:a0a1}.

  To
quantify the uncertainty related to the extrapolation separately from the uncertainty due  to  the
variation of the Borel parameter $M^2$ and of the continuum threshold $s_0$, we
considered the range $[q^2_L, q^2_H]$  previously used to determine the fitting parameters $a_0$ and $a_1$. Since the sum rule is
more reliable for large negative values of $q^2$, we shifted the lower boundary $q^2_L$ by $\pm 2$\ GeV$^2$ and the upper boundary
$q^2_H$ by $\pm 1$\ GeV$^2$, generating  five intervals. For each of the five
resulting fits we computed the extrapolated form factor at zero recoil and took $\delta k_{extrap}$ as half the spread between
the maximum and minimum extrapolated values. We then combined it in quadrature with the uncertainty $\delta k_{method}$
obtained from the variation of $M^2$ and $s_0$. In the worst case, 
the total uncertainty increases by less than 4\%. This shows that the choice of fitting range is not a source of
significant uncertainty compared to LCSR input parameters.

\begin{figure}[h]
\begin{center}
\includegraphics[width = 0.7\textwidth]{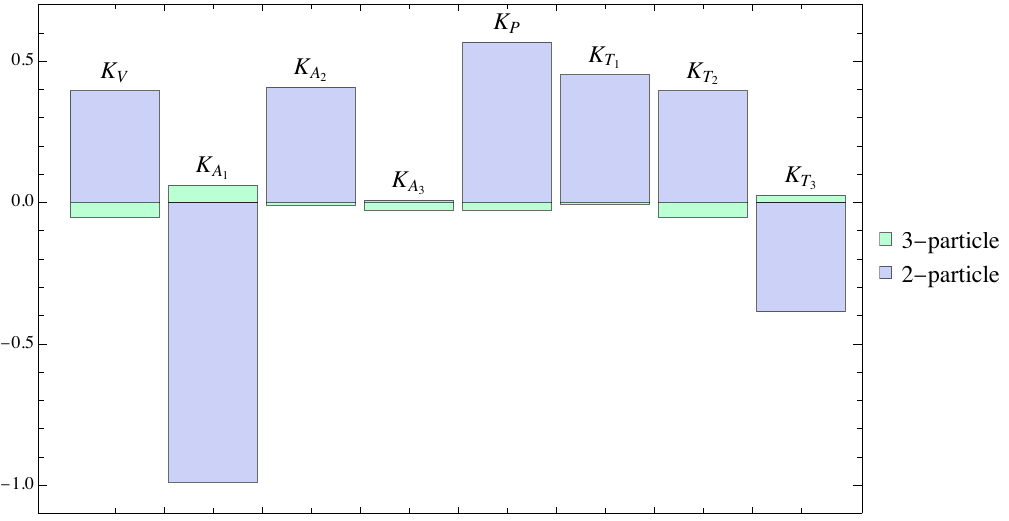} 
\caption{\baselineskip 10pt  \small  Graphical representation of the numerical impact of the  two- and three-particle contributions  to the sum rule results for the various form factors at $q^2=0$.
}\label{fig:3vs2}
\end{center}
\end{figure}

It is instructive to quantify the numerical impact of the three-particle contributions in the case of the $B \to D_2^*$ transition. This is illustrated in Fig.~\ref{fig:3vs2}, where the LCSR predictions obtained with and without the three-particle contributions are compared for all form factors at $q^2=0$. As can be seen, the size of the three-particle contributions depends significantly on the form factor. They amount to about $15\%$ for $k_V$ and $k_{T_2}$, to about $5$-$7\%$ for $k_{A_1}$, $k_P$, and $k_{T_3}$, and to about $1$-$3\%$ for $k_{A_2}$ and $k_{T_1}$. A much larger effect, of about $140\%$, is found for $k_{A_3}$, owing to the opposite signs of the two- and three-particle contributions. This large relative correction is due to the accidental suppression of the two-particle contribution and therefore does not imply an equally large effect on physical observables.
 This is addressed in the next section, where the  impact on the branching fractions and on the ratio $R(D_2^*)$ is discussed.

%


\begin{figure}[t]
\begin{center}
\includegraphics[width = 0.38\textwidth]{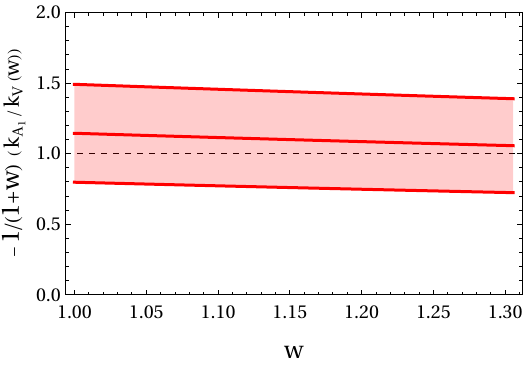} \hskip 0.3cm 
\includegraphics[width = 0.38\textwidth]{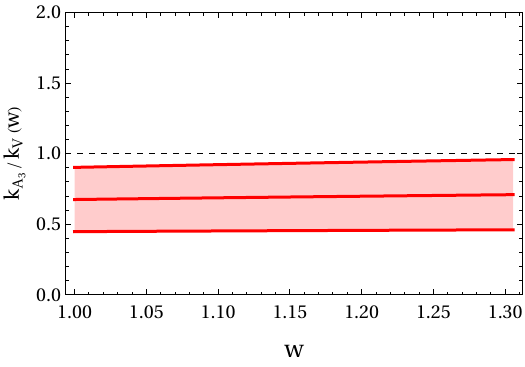} \\
\includegraphics[width = 0.38\textwidth]{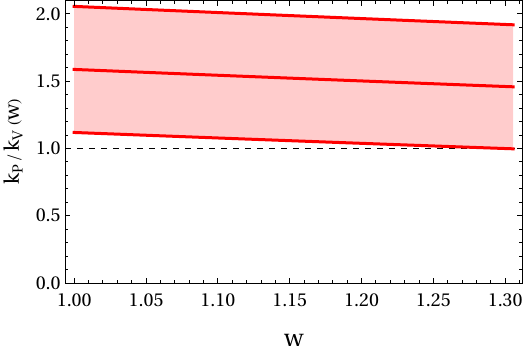} \hskip 0.3cm
\includegraphics[width = 0.38\textwidth]{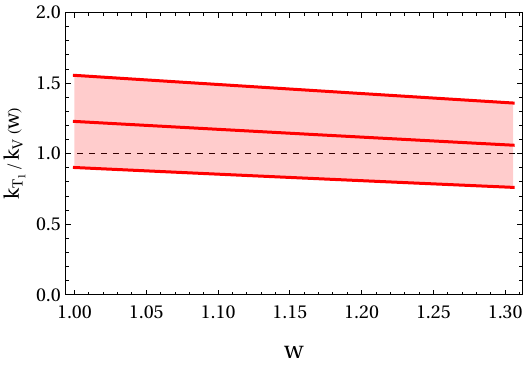} \\
\includegraphics[width = 0.38\textwidth]{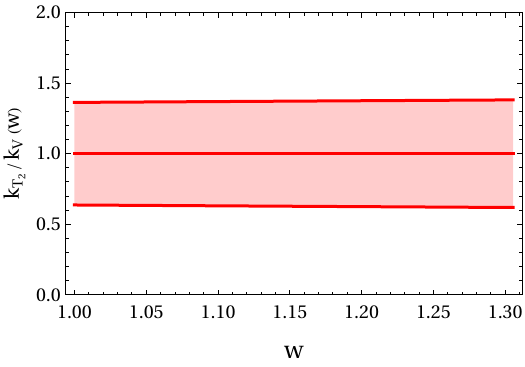} 
\caption{\baselineskip 10pt  \small   Ratios in eq.~\eqref{RFV}. They are  expected to be 1 in the HQ limit.
}\label{fig:HQ1}
\end{center}
\end{figure}
\begin{figure}[b]
\begin{center}
\includegraphics[width = 0.38\textwidth]{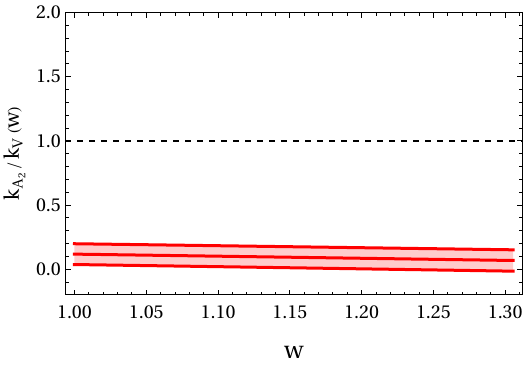} \hskip 0.3cm 
\includegraphics[width = 0.38\textwidth]{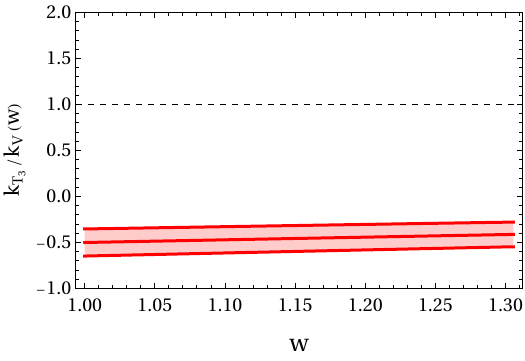}
\caption{\baselineskip 10pt  \small   Ratios in eq.~\eqref{RFV}. They are  expected to vanish in the HQ limit.
}\label{fig:HQ2}
\end{center}
\end{figure}

In the heavy quark limit $m_Q \to \infty$  the form factors obey the relations
\be
k_V({w})=-\frac{1}{(1+{w})}k_{A_1}({w})=k_{A_3}({w})=k_P({w})=k_{T_1}({w})=k_{T_2}({w})=\tau_{3/2}({w}) \label{HQrel}
\ee
and 
\be
k_{A_2}({w})=k_{T_3}({ w})=0 \,\,\, , \label{HQrel1}
\ee
where ${w}=(m_B^2-m_{D^*_2}^2-q^2)/(2m_B m_{D^*_2})$. $\tau_{3/2}({w})$ is the universal Isgur–Wise function describing the transitions from the heavy-light  S-wave $J^P_{j_\ell}=(0^-,\,1^-)_{1/2}$ meson doublet  to the P-wave $J^P_{j_\ell}=(1^+,\,2^+)_{3/2}$ doublet,  $j_\ell$ being the angular momentum of the light degrees of freedom.
In figs.~\ref{fig:HQ1} and \ref{fig:HQ2} we plot the ratios
\be
R_{FV}({w})=g_F({w})\frac{k_F({w})}{k_V({w})} \label{RFV} 
\ee
for the   computed $B \to D_2^*$  form factors $k_F(w)$. 
The function $g_F({w})$ is   $g_{A_1}({ w})=-1/(1+{ w})$ for $k_{A_1}$, and  $g_F({ w})=1$ for all other form factors. The  bands in the figures correspond to the errors quoted in table~\ref{tab:a0a1}.
In the HQ limit the ratios plotted  in fig.~\ref{fig:HQ1} are expected to be 1:  we find that $R_{A_1V}$, $R_{T_1V}$, and $R_{T_2V}$  satisfy this expectation within the uncertainties, while sizable deviations affect $R_{PV}$ and $R_{A_3V}$.
The ratios  in fig.~\ref{fig:HQ2} are expected to vanish in the HQ limit: this behavior is fulfilled for $R_{A_2V}$ better than for $R_{T_3V}$. All  results provide a hint on the size of the heavy quark mass corrections for the various form factors.

\section{Phenomenological implications,  comparison with previous analyses}\label{sec:pheno}
Using the parametrization in \eqref{kFfit} and the values in table \ref{tab:a0a1} we can compute the $q^2$ spectra and the decay rates of  $B_q \to H^*_2 \ell \bar \nu_\ell$ in the Standard Model.
We define the dimensionless variables
\bea
{\hat q}^2=\frac{q^2}{m_{B_q}^2} \,, \hskip 1 cm r=\frac{m_{H_2^*}}{m_{B_q}} \,, \hskip 1 cm \rho_L=\frac{m_\ell}{m_{B_q}}\,\,,
\eea
and 
\bea
p_\pm&=& {\hat \lambda}^{1/2} \,k_V \pm 2 r \, k_{A_1} \label{ppiumeno}
\\
p_0&=& r\, k_{A_2}(1+{\hat q}^2-r^2)+k_{A_3}(1-{\hat q}^2-r^2)+2r\, k_{A_1}
\label{p0}
\\
p_1&=& (r\, k_{A_2}+k_{A_3}){\hat \lambda}+2r\, k_{A_1}(1-{\hat q}^2-r^2)
\label{p1}
\eea
where ${\hat \lambda}=\lambda(1,\,r^2,\,{\hat q}^2)$ and $\lambda$  the K\"all\'en function.
The ${\hat q}^2$ distribution reads
\bea
\frac{d\Gamma (B_q \to H^*_2 \ell \bar \nu_\ell)}{d {\hat q}^2}&=&\Gamma_0 \frac{{\hat \lambda}^{3/2}}{192 \,{\hat q}^6 r^5} \big({\hat q}^2-\rho_L^2 \big)^2
\nn \\
&\times&\Big\{  \big(2{\hat q}^2+\rho_L^2 \big) \Big[3 {\hat q}^2  r^2(p_+^2+p_-^2)+p_1^2 \Big] +3 {\hat \lambda}\rho_L^2 p_0^2 \Big\} \label{rate} 
\eea
where $\Gamma_0=\displaystyle\frac{G_F^2 |V_{cb}|^2 m_{B_q}^5}{192 \pi^3}$.
Using the values of meson masses and lifetimes quoted in \cite{ParticleDataGroup:2024cfk}
 we obtain
\bea
{\cal B}(B^- \to D_2^{*0} \,\mu^- {\bar \nu}_\mu)&=&(3.3 \pm 1.3) \times 10^{-3} \, \left(\frac{|V_{cb}|}{0.04}\right)^2
\nn \\
{\cal {\bar B}}(B^0_s \to D_{s2}^{*+} \,\mu^- {\bar \nu}_\mu)&=& ( 6.0 \pm 1.8) \times 10^{-3} \, \left(\frac{|V_{cb}|}{0.04}\right)^2 \,\,
\label{brs}
\eea
and 
\bea
{\cal B}(B^- \to D_2^{*0} \,\tau^- {\bar \nu}_\tau)&=&(1.8 \pm 0.7) \times 10^{-4} \, \left(\frac{|V_{cb}|}{0.04}\right)^2
\nn \\
{\cal {\bar B}}(B^0_s \to D_{s2}^{*+} \,\tau^- {\bar \nu}_\tau)&=& ( 3.3\pm 0.9 ) \times 10^{-4} \, \left(\frac{|V_{cb}|}{0.04}\right)^2 \,\,.
\label{brs1}
\eea
For $B_s$,  the notation ${\cal {\bar B}}$  indicates that   the $B_s - {\bar B}_s$ mixing  is 
taken into account by the factor $1/(1-y_s)$  with $y_s=0.061 \pm 0.009$ \cite{LHCb:2014iah}.
Measurements  are   available for  the muon mode \cite{ParticleDataGroup:2024cfk}:
\bea
{\cal B}(B^- \to D_2^{*0} \,\mu^- {\bar \nu}_\mu)\,{\cal B}(D_2^{*0} \to D^+ \pi^-)&=&(1.59 \pm 0.10) \times 10^{-3} \nn \\
{\cal B}(B^- \to D_2^{*0} \,\mu^- {\bar \nu}_\mu)\,{\cal B}(D_2^{*0} \to D^{*+} \pi^-)&=&(1.06 \pm 0.18) \times 10^{-3}  
\label{brexp}\\
{\cal {\bar B}}(B^0_s \to D_{s2}^{*+} \, X\,\mu^- {\bar \nu}_\mu)\,{\cal B}(D_{s2}^{*+} \to {\bar D}^0 K^+)&=&(2.7\pm 1.0) \times 10^{-3} \nn \,\, .
\eea
For the $\tau$  modes,  the LHCb Collaboration reports evidence for $B^- \to D^{**0} \,\tau^- \bar{\nu}_\tau$, with $D^{**0}$  the set of excited $P$-wave charmed mesons \cite{LHCb:2025fri}.

The  anomalies observed in the LFU  ratios ${\cal R}(D^{(*)})$  
motivate the investigation of analogous observables in $\bar B \to H^*_2$. 
Besides providing additional tests of lepton flavour universality, such quantities also affect the experimental determination of ${\cal R}(D^{(*)})$,  the $B$ to tensor meson modes representing important backgrounds \cite{Bernlochner:2021vlv}.
We find 
\bea
{\cal R}(D_2^*)&=&\frac{{\cal B}(B^- \to D_2^{*0} \,\tau^- {\bar \nu}_\tau)}{{\cal B}(B^- \to D_2^{*0} \,\mu^- {\bar \nu}_\mu)} =0.055 \pm 0.003 
\label{RD2}\\
{\cal R}(D_{s2}^*)&=&\frac{{\cal {\bar B}}(B^0_s \to D_{s2}^{*+} \,\tau^- {\bar \nu}_\tau)}{{\cal {\bar B}}(B^0_s \to D_{s2}^{*+} \,\mu^- {\bar \nu}_\mu)}=0.054 \pm 0.003 \,\,.
\label{RDs2}
\eea
These results are consistent with the ones in \cite{Biancofiore:2013ki} based on the $\tau_{3/2}$ obtained by  three-point QCD sum rules in the heavy quark limit  \cite{Colangelo:1992kc}.

 To quantify the impact of the three-particle contributions on the physical observables, we have repeated the calculation of the branching fractions and of the ratio ${\cal R}(D_2^*)$ by retaining only the two-particle contributions in the form factors. We find
that  the inclusion of the three-particle contributions increases the branching fractions by about $12\%$ and $17\%$ for the $\mu$ and $\tau$ modes, respectively, whereas their effect on the ratio ${\cal R}(D_2^*)$ is much smaller, amounting to about $2\%$. This behavior reflects the partial cancellation of common effects in the ratio.

The outcome of our  calculation  can be compared with previous findings. The $\bar B \to H^*_2$ form factors  have been  investigated   in the context of  determining the $\tau_{3/2}({w})$ universal function. Previous literature has employed  short-distance QCD sum rules for this purpose \cite{Colangelo:1992kc, Dai:1998ca, Huang:2001qh,Zuo:2023ksq}.
In \cite{Colangelo:1992kc}, the universal function was extracted from the vector form factor $k_V$. After accounting for the different normalization, the value $\tau_{3/2}(1) \simeq 0.374$ was obtained. In contrast,  $\tau_{3/2}(1) = 0.74 \pm 0.15$ was reported in \cite{Dai:1998ca}. For comparison, our present calculation yields $k_V(1) = 0.503 \pm 0.065$.
 Next-to-leading order corrections to the HQ limit were addressed  in \cite{Huang:2001qh,Zuo:2023ksq}.
 Bakamjian--Thomas quark models  yielded $\tau_{3/2}(1) \simeq 0.5$  \cite{Morenas:1997nk,LeYaouanc:2021xcq}. In \cite{Bernlochner:2016bci}, form factors were extracted at next-to-leading order in the HQ expansion  via fits to experimental data; however,  a comparison with our results would require an update of the experimental input used therein.
In full QCD  the form factors  have been computed in \cite{Azizi:2013aua,Aliev:2019ojc}.  The short-distance sum rules employed in \cite{Azizi:2013aua} give 
branching fractions that  deviate significantly from  the results presented here.
The LCSR formalism with external $B$ state  was also utilized  in \cite{Wang:2010tz,Aliev:2019ojc}.  With respect to  \cite{Aliev:2019ojc}  we add the three-particle contributions,   we compute   the 
$B_s \to D_{s2}^*$  form factors and we  determine  $k_P$.   A more detailed comparison with Ref.~\cite{Aliev:2019ojc} requires taking into account the different conventions adopted in the two analyses. Ref.~\cite{Aliev:2019ojc} employs a different definition of the $D_2^*$ decay constant and a different interpolating current for the tensor meson. Moreover, the form factors are defined according to a different parametrization. After translating the form-factor definitions of Ref.~\cite{Aliev:2019ojc} into our conventions and retaining only the two-particle contributions, we find that all corresponding sum rules agree up to an overall factor $4m_{D_2^*}$.
The factor $m_{D_2^*}$ originates from the different normalization adopted for the tensor-meson decay constant, whereas the remaining factor $4$ is  due to the different choice of the interpolating current. This  relation is satisfied for all form factors except for $K_{T_1}$ and $K_{T_2}$, in the terms proportional to $\overline{G}_{\pm}$.

Numerical discrepancies also arise from the different choices of continuum subtraction and Borel parameters, as well as the use of  $f_{D_2^*}$  from \cite{Wang:2014yza},   which differs from our determination.

\section{Conclusions  and outlook}\label{sec:Conclusion}

We  summarize the new results  of this work, pointing out  their phenomenological relevance.

\begin{itemize}
\item We have presented a  new determination of the  vacuum-particle matrix element of a quark current interpolating  non-strange and strange charmed and beauty mesons with $J^P=2^+$. The calculation  improves the results available in the literature by the correct inclusion of the $D=4$ contribution, giving the prescription for the regularisation of the threshold singularity in the case of the nonvanishing strange quark mass. The impact of previously neglected effects has been studied in detail, namely the inclusion of the finite width of the resonance in the sum rule.

\item We have computed  the form factors parametrizing the $\bar B \to H^*_2$ matrix elements of SM quark currents,  using LCSR with external $B$ state.
With respect to previous analyses, we have included the three-particle contributions and studied their impact in detail. This is a step forward in the systematic determination of the theoretical accuracy. In the numerical analyses  we have employed our computed  decay constants. We have investigated the effect of the extrapolation in the squared momentum transfer from the negative values, where the sum rules are more reliable, to the physical kinematic range,  and we have determined  the related errors in the form factor parameters.
\item Using the same approach, we have determined the  form factors of the  $\bar B \to H^*_2$ matrix elements involving pseudoscalar and tensor quark currents, as featured in the generalized  low-energy  Hamiltonian describing the  $b \to c \ell \bar \nu_{\ell}$ transition. These results are important for the precision analyses of the upcoming experimental measurements in model-independent approaches based on the effective field theory. they will  enable us to constrain  NP Wilson coefficients,  as in  the $B \to D^*(D \pi) \ell \bar \nu_\ell$ case \cite{Colangelo:2024mxe}.
\item We have computed the semileptonic branching fractions and LFU ratios within the SM, with carefully identified uncertainties.
\item We have tested the heavy quark limit for the various form factors,  identifying the cases where  the heavy quark mass corrections are large.
\end{itemize}

The results can be improved in several aspects, namely by incorporating NLO QCD perturbative corrections into the two-point spectral density: such refinements are deferred to future calculations.
In the meanwhile, they can be used for the phenomenological analyses and are helpful for designing the new measurements at present and future facilities.
\section*{Acknowledgments}
This work has  been carried out within the project (Iniziativa Specifica) "Precision Studies for Fundamental Interactions" (SPIF) of Istituto Nazionale di Fisica Nucleare.
%
\newpage
\appendix 
\numberwithin{equation}{section}
\section{Sum rule for  $f_{H^*_2}$}\label{appA}
We collect  the expressions of the OPE terms  in eq.~\eqref{srborel}.
\begin{itemize}
\item  Perturbative spectral density:
\be
\rho^{\rm pert}(s)=\frac{1}{2 \pi^2 s^3} \left[s-(m_Q-m_q)^2 \right] \left[2 (m_Q +  m_q)^2 + 3 s\right] \lambda^{3/2}(s,m_Q^2,m_q^2) \label{pert2pt}
\ee
with $\lambda$  the K\"all\'en function.

\item  $D=3$:
\be
\widehat \Pi^{\rm QCD,\,D=3}(M^2) = \langle{\bar q}q \rangle\,45 m_Q m_q^2\,e^{-m_Q^2/M^2} .
\ee

\item  $D=4$:
The OPE spectral density proportional to the gluon condensate consists of  five contributions, three derived expanding the quark propagators, and two originating from a gluon  emitted from the vertex with the covariant derivative:
\par
\bea
\rho^{D=4,(1)}(s)&=&\frac{5m_Q}{12 \pi^2 s^3 \left[s- (m_Q +  m_q)^2 \right]  \lambda^{1/2}(s,m_Q^2,m_q^2) } \nn \\
 &\times& \Big\{ 3 m_q\left[s- (m_Q +  m_q)^2 \right]^4 \nn \\
&-&\left[s- (m_Q +  m_q)^2 \right]^3 (2m_Q^3-2m_q m_Q^2-19 m_q^2 m_Q-9 m_q^3)\nn \\
&+&3m_q\left[s- (m_Q +  m_q)^2 \right]^2(-3m_Q^4+m_q m_Q^3+13 m_q^2 m_Q^2+13 m_q^3 m_Q+4 m_q^4)\nn \\
&+&6 m_q^2(m_q+m_Q)^3\left[s- (m_Q +  m_q)^2 \right](m_q^2+2 m_q m_Q -m_Q^2) \nn \\
&+&2m_q^3 m_Q(m_Q +  m_q)^5 \Big\} \label{g21}
\eea
\bea
\rho^{D=4,(2)}(s)&=&\frac{5m_q}{12 \pi^2 s^3 \left[s- (m_Q +  m_q)^2 \right]  \lambda^{1/2}(s,m_Q^2,m_q^2) } \nn \\
&\times& \Big\{3 m_Q\left[s- (m_Q +  m_q)^2 \right]^4 \nn \\
&-&\left[s- (m_Q +  m_q)^2 \right]^3 (2m_q^3-2m_q^2 m_Q-19 m_q m_Q^2-9 m_Q^3)\nn \\
&+&3m_Q\left[s- (m_Q +  m_q)^2 \right]^2(-3m_q^4+m_Q m_q^3+13 m_q^2 m_Q^2+13 m_q m_Q^3+4 m_Q^4)\nn \\
&+&6 m_Q^2(m_q+m_Q)^3\left[s- (m_Q +  m_q)^2 \right](-m_q^2+2 m_q m_Q +m_Q^2) \nn \\
&+&2m_Q^3 m_q(m_Q +  m_q)^5\Big\} \label{g22}
\eea
\newpage
\bea
\rho^{D=4,(3)}(s)&=&\frac{5}{72 \pi^2 s^5  \lambda^{1/2}(s,m_Q^2,m_q^2) }\nn \\
&\times& \Big\{-5\left[s- (m_Q +  m_q)^2 \right]^6\nn \\
&+&5\left[s- (m_Q +  m_q)^2 \right]^5 \Big(m_q^2+m_Q^2-6(m_q+m_Q)^2 \Big)\nn \\
&+&\left[s- (m_Q +  m_q)^2 \right]^4\Big(9 m_q^4 + 10 m_q^2 m_Q^2 + 9 m_Q^4 - 75 (m_q + m_Q)^4 \nn \\
&+& 25 (m_q + m_Q)^2 (m_q^2 + m_Q^2)\Big) \nn \\
&-&\left[s- (m_Q +  m_q)^2 \right]^3\Big(21 m_q^6 + 328 m_q^5 m_Q + 1061 m_q^4 m_Q^2+ 1520 m_q^3 m_Q^3  \nn \\
&+& 1061 m_q^2 m_Q^4 + 328 m_q m_Q^5 + 21 m_Q^6\Big) \nn \\
&-&2 (m_q + m_Q)^2 \left[s- (m_Q +  m_q)^2 \right]^2 \Big(3 m_q^6 + 57 m_q^5 m_Q + 331 m_q^4 m_Q^2 \nn \\ 
&+& 464 m_q^3 m_Q^3  + 331 m_q^2 m_Q^4 + 57 m_q m_Q^5 + 3 m_Q^6\Big) \nn \\
&-&6 m_q m_Q (m_q + m_Q)^4 \left[s- (m_Q +  m_q)^2 \right] \Big(4 m_q^4 + 19 m_q^3 m_Q \nn \\
&+& 60 m_q^2 m_Q^2 + 19 m_q m_Q^3 + 4 m_Q^4\Big) \nn \\
&-&12 m_q^2 m_Q^2 (m_q + m_Q)^6 (m_q^2 + m_q m_Q + m_Q^2)\Big\} \label{g23}
\eea
\bea
\rho^{D=4,(4)}(s)&=&\frac{5}{48 \pi^2 s^2  \lambda^{1/2}(s,m_Q^2,m_q^2) } \nn \\
&\times&
 \Big\{\left[s- (m_Q +  m_q)^2 \right]^3 + 2\left[s- (m_Q +  m_q)^2 \right]^2 \left(m_q + m_Q\right)^2 \nn \\
&+&  \left[s- (m_Q +  m_q)^2 \right] \left(2 m_Q (m_Q^3 + 3 m_q^2 m_Q + 3 m_q m_Q^2 + 5 m_q^3) \right) \nn \\ 
&+&8m_q^2 m_Q^2(m_Q+m_q)^2 \Big\} \label{g24}
\eea
 \bea
\rho^{D=4,(5)}(s)&=&\frac{5}{48 \pi^2 s^2  \lambda^{1/2}(s,m_Q^2,m_q^2) } \nn \\
 &\times& \Big\{\left[s- (m_Q +  m_q)^2 \right]^3 + 2\left[s- (m_Q +  m_q)^2 \right]^2 \left(m_q + m_Q\right)^2 \nn \\
&+&  \left[s- (m_Q +  m_q)^2 \right] \left(2 m_q (m_q^3 + 3 m_q^2 m_Q + 3 m_q m_Q^2 + 5 m_Q^3) \right) \nn \\ 
&+&8m_q^2 m_Q^2(m_Q+m_q)^2 \Big\} \,\, . \label{g25}
\eea
The  resulting $D=4$ contribution in the OPE side of the sum rule reads
\be
\rho^{D=4}(s)= \langle \frac{\alpha_s}{\pi} G^2 \rangle  (2 \pi)^2 \sum_{i=1}^5 \rho^{D=4,(i)}(s) .\label{rhoG2}
\ee
\item $D=5$: 

\be
\widehat \Pi^{\rm QCD,\,D=5}(M^2) =
 -\frac{70}{3}\langle{\bar q}g_s\sigma \cdot G q\rangle\,m_Q \,e^{-m_Q^2/M^2}\,\,.
\ee

\end{itemize}

\section{LCSR for the form factors}\label{appB}
We collect the expressions appearing in eqs.~\eqref{sr-gen}, \eqref{sr-gen1}, \eqref{sr-gen2} and \eqref{sr-gen3}, needed to compute  the  form factors  parametrizing the  $B_q \to H^*_2$ matrix elements in  eq.~\eqref{ff:chic2}.
To simplify the notation, we omit the argument of the LCDA distribution functions:  the functions entering in the two-particle contribution  $\phi_+$, $g_+$, ${\bar \phi}_\pm$ and $\bar G_\pm$  only depend  on $\sigma$, the functions entering in the three-particle contribution $\psi_A$, $\psi_V$, ${\bar X}_A$, ${\tilde {\bar X}}_A$, ${\bar Y}_A$, ${\tilde {\bar Y}}_A$, ${\bar {\bar W}}$ and ${\bar {\bar Z}}$ depend on $(\omega_1,\omega_2)$.
\begin{itemize}
\item{$k_V$}
\bea
c^{had}_V &=& -\frac{1}{2}f_{H^*_2} m_{H^*_2}^2 \sqrt{m_B m_{H^*_2}} \,\, , \quad
c^{OPE}_V =2 f_B m_B^3 m_{H^*_2} \,\, , \quad
c_V=-\frac{4 f_B m_B^{5/2}}{f_{H^*_2} m_{H^*_2}^{3/2}} \,\, , \qquad
\eea
\bea
{\bar{ \cal F}}^{2pt(V)}_1 (\sigma)&=&\frac{\sigma}{\bar \sigma}\phi_+  \\
{\bar{ \cal F}}^{2pt(V)}_2 (\sigma)&=&\frac{\sigma}{{\bar \sigma}^2}\big[8g_+ +m_c \,{\bar \phi}_\pm \big] \\
{\bar{ \cal F}}^{2pt(V)}_3 (\sigma)&=&8m_c\frac{\sigma}{{\bar \sigma}^3}\big[{\bar G}_\pm -m_c \,g_+ \big]  \\
{\bar{ \cal F}}^{2pt(V)}_4 (\sigma)&=&-24 m_c^3 \,\frac{\sigma}{{\bar \sigma}^4}\, {\bar G}_\pm  
\eea
\bea
{\bar{ \cal P}}^{3pt(V)}_1 (\sigma,\omega_1,\omega_2)&=&0 \\
{\bar{ \cal P}}^{3pt(V)}_2 (\sigma,\omega_1,\omega_2)&=&\frac{\sigma}{\sigmabar^2}\Big[\frac{(2 \omega_1 + \omega_2 - 2 m_B \sigma)}{\omega_2}\psi_A+\psi_V\Big] \nn \\
&+&\frac{1}{m_B \omega_2\,\sigmabar^3}\Big\{\Big[4\sigma (m_B \sigma-\omega_1)-\omega_2(1+\sigma)\Big]{\bar X}_A\nn \\
&+&\Big[\omega_2(1+\sigma)-2\sigmabar \big(m_B \sigma -\omega_1\big) \Big] {\tilde {\bar X}}_A\Big\}  \\
{\bar{ \cal P}}^{3pt(V)}_3 (\sigma,\omega_1,\omega_2)&=&\frac{2\sigma}{m_B\,\omega_2 \, \sigmabar^4}\Big\{-\omega_2\big[(m_B \sigmabar+m_c)^2-q^2 \big]\nn \\
&+&2(m_B\sigma -\omega_1)\big[m_B^2\sigmabar^2+m_c^2-q^2\big]
\Big\}\,{\bar X}_A\nn \\
&+&2\frac{\sigma}{m_B\, \sigmabar^4}\big[(m_B \sigmabar-m_c)^2-q^2 \big]{\tilde{\bar X}}_A \nn \\
&+&4\frac{(m_B \sigma-\omega_1)}{m_B \, \omega_2^2 \,\sigmabar^4}\Big[m_c\omega_2(2\sigma +1)+2m_B \sigma\,\sigmabar(m_B \sigma-\omega_1-\omega_2) \Big]{\bar{\bar W}}\nn \\
&+&4\frac{\sigma}{\sigmabar^3}\Big\{m_c\big[{\bar W}+{\bar Y}_A+{\tilde{\bar Y}}_A\big]+2\frac{(m_B \sigma -\omega_1)}{\omega_2}{\bar{\bar Z}}\Big\}  \\
{\bar{ \cal P}}^{3pt(V)}_4 (\sigma,\omega_1,\omega_2)&=&24m_c^2\, \frac{\sigma}{\sigmabar^4}\frac{(m_B \sigma -\omega_1)(2m_B \sigma-2\omega_1-\omega_2)}{\omega_2^2}{\bar{\bar Z}}\nn \\
&+&12m_c \, \frac{\sigma \, (m_B \sigma-\omega_1)}{m_B \, \omega_2^2 \,\sigmabar^5}
\Big\{\omega_2\big[(m_B \sigmabar+m_c)^2-q^2 \big] \nn \\
&-&4m_B m_c \sigmabar (m_B \sigma-\omega_1)\Big\}{\bar{\bar W}} 
\eea
\bea
{\bar{ \cal R}}^{3pt(V)}_1 (\sigma,\omega_1,\omega_2)&=&0 \\
{\bar{ \cal R}}^{3pt(V)}_2 (\sigma,\omega_1,\omega_2)&=&2\frac{m_B\sigma-\omega_1}{m_B \omega_2 \sigmabar^3}
\Big[-\sigmabar \big[m_c \psi_A-(m_c+m_B \sigma)\psi_V\big]+(-1+3\sigma){\bar X}_A \Big]  \nn \\ \\
{\bar{ \cal R}}^{3pt(V)}_3 (\sigma,\omega_1,\omega_2)&=&4\frac{m_B\sigma-\omega_1}{m_B \omega_2 \sigmabar^4}
\Big\{ \Big[\sigma \big(m_B^2 \sigmabar^2-q^2-m_c^2 \big) +2m_c^2 \Big]{\bar X}_A \nn \\
&-&2m_c m_B \sigma\sigmabar \big[{\bar W}+{\bar Y}_A \big]\nn \\
&+&2 \frac{(m_B \sigma -\omega_1)}{\omega_2}\big[  -m_c (1+2 \sigma){\bar{\bar W}}+2m_B \sigma \sigmabar {\bar{\bar Z}}  \big] \Big\}  \\
{\bar{ \cal R}}^{3pt(V)}_4 (\sigma,\omega_1,\omega_2)&=&24m_c\frac{(m_B\sigma-\omega_1)^2}{m_B \omega_2^2 \sigmabar^5}\Big\{\Big[-\sigma \big(m_B^2 \sigmabar^2-q^2-m_c^2 \big) -2m_c^2 \Big]{\bar{\bar W}}\nn \\
&-&2 m_B m_c\sigma \sigmabar {\bar{\bar Z}}   \Big\}
\eea
\vskip 0.5cm
\item{$k_{A_1}$}
\bea
c^{had}_{A_1} &=& -\frac{i}{2}f_{H^*_2} m_{H^*_2}^2 \sqrt{m_B m_{H^*_2}}  \,\, , \quad
c^{OPE}_{A_1} =i f_B m_B  \,\, , \quad
c_{A_1}=-\frac{2 f_B \sqrt{m_B}}{f_{H^*_2} m_{H^*_2}^{5/2}} \,\, , \qquad
\eea
\bea
{\bar{ \cal F}}^{2pt(A_1)}_1 (\sigma)&=&-m_B\frac{\sigma}{\bar \sigma^2}\Big\{ \big[(m_B \sigmabar+m_c)^2-q^2 \big]\phi_++8 g_+ +m_c {\bar \phi}_\pm  \Big\}  \\
{\bar{ \cal F}}^{2pt(A_1)}_2 (\sigma)&=&-m_B\frac{\sigma}{{\bar \sigma}^3}\Big\{8\big[m_B \sigmabar(m_B \sigmabar+m_c)-q^2\big]g_++8(2m_B \sigmabar +m_c){\bar G}_\pm \nn  \\
&+& m_c\big[(m_B \sigmabar+m_c)^2-q^2 \big]{\bar \phi}_\pm \Big\} \\
{\bar{ \cal F}}^{2pt(A_1)}_3 (\sigma)&=&8m_Bm_c\frac{\sigma}{{\bar \sigma}^4} \Big\{m_c\big[(m_B \sigmabar+m_c)^2-q^2 \big]g_+ -\big[m_B^2 \sigmabar^2-2m_c^2-q^2 \big] {\bar G}_\pm \Big\} \nn \\ \\
{\bar{ \cal F}}^{2pt(A_1)}_4 (\sigma)&=&24m_B m_c^3\frac{\sigma}{{\bar \sigma}^5}\big[(m_B \sigmabar+m_c)^2-q^2 \big]{\bar G}_\pm 
\eea
\bea
{\bar{ \cal P}}^{3pt(A_1)}_1 (\sigma,\omega_1,\omega_2)&=&\frac{m_B \sigma}{\sigmabar^2}\Big[-\psi_V+\frac{2m_B \sigma -2\omega_1 -\omega_2}{\omega_2}\psi_A\ \Big]\nn \\
&+&\frac{1}{\omega_2 \sigmabar^3} \Big\{ \big[\omega_2 (1+\sigma)+4 \sigma (\omega_1-m_B\sigma) \big]\,{\bar X}_A  \nn \\
&+& \big[2\sigmabar(m_B \sigma -\omega_1)-\omega_2(1+\sigma) \big]\,{\tilde {\bar X}}_A \Big\}  
\eea
\bea
{\bar{ \cal P}}^{3pt(A_1)}_2 (\sigma,\omega_1,\omega_2)&=&\frac{m_B \sigma}{\omega_2 \sigmabar^3} \Big\{\omega_2\big[(m_B \sigmabar+m_c)^2-q^2 \big]\nn \\
&+&2(m_B\sigma -\omega_1-\omega_2)\big[m_B^2\sigmabar^2+m_c^2-q^2\big]\Big\}\psi_A \nn\\
&-&\frac{m_B \sigma}{ \sigmabar^3}\big[(m_B \sigmabar+m_c)^2-q^2 \big]\psi_V  \nn \\
&+&\frac{1}{\omega_2 \sigmabar^4}\Big\{8 m_B (-m_c^2 + q^2 - m_B^2 \sigmabar^2) \sigma^2+2 m_B m_c \omega_2\sigmabar^2 \nn \\
&+&(m_c^2 - q^2) (\omega_2 + 8 \omega_1 \sigma + 3 \omega_2 \sigma) +  m_B^2 \sigmabar^2 (\omega_2 + 8 \omega_1 \sigma + 5 \omega_2 \sigma)\Big\}{\bar X}_A  \nn \\
 &+&\frac{1}{\omega_2 \sigmabar^4}\Big\{ 2 m_B (m_c^2 - q^2 + m_B^2(1-\sigma^2) ) \sigma \sigmabar  +2 m_B m_c \omega_2\sigmabar^2 \nn \\
 &-&(m_c^2 - q^2) (2\omega_1  \sigmabar+  \omega_2 (1+3\sigma)) \nn \\
 &-&m_B^2 \sigmabar^2 (\omega_2 + 2 \omega_1 (1+\sigma) + 5 \omega_2 \sigma)\Big\}{\tilde {\bar X}}_A\nn \\
 &-&\frac{4m_B \sigma}{\omega_2 \sigmabar^3}\big[m_c \omega_2 + m_B (4 \omega_1 + 3 \omega_2) \sigmabar - 4 m_B^2 \sigmabar \sigma \big]\big({\bar Y}_A+{\bar W} \big)\nn \\
&-&\frac{4m_B \sigma}{\omega_2 \sigmabar^3}\big[m_c \omega_2 - m_B (2 \omega_1 + 3 \omega_2) \sigmabar +2m_B^2 \sigmabar \sigma \big]{\tilde {\bar Y}}_A\nn \\
&-&\frac{4(m_B \sigma-\omega_1)}{\omega_2^2 \sigmabar^4}\Big[-6 m_B^2 \sigmabar \sigma^2 + m_c \omega_2 (1 + 2 \sigma) \nn \\ 
&+&m_B \sigmabar(\omega_2 + 6 \omega_1 \sigma + 3 \omega_2 \sigma)\Big]{\bar {\bar W}} +\frac{8m_B \sigma}{\omega_2 \sigmabar^3}(\omega_1-m_B \sigma){\bar {\bar Z}}
\eea
\bea
&&{\bar{ \cal P}}^{3pt(A_1)}_3(\sigma,\omega_1,\omega_2)=-\frac{4m_B m_c \sigma}{\omega_2 \sigmabar^4}\Big\{
\omega_2 \big[(m_B \sigmabar+m_c)^2-q^2]{\tilde {\bar Y}}_A  \nn \\
&+& \Big[4m_B m_c \sigmabar(m_B \sigma -\omega_1)+\omega_2 \big[(m_B \sigmabar-m_c)^2-q^2]\Big]\Big({\bar W}+{\bar Y}_A \Big) \Big\}\nn \\
&-&\frac{2\sigma}{\omega_2 \sigmabar^5}  \big[(m_B \sigmabar+m_c)^2-q^2\big]\big[(m_B \sigmabar-m_c)^2-q^2\big]\Big[(2m_B \sigma-2\omega_1-\omega_2){\bar X}_A+{\tilde {\bar X}}_A \Big]\nn \\
&+&\frac{4(m_B \sigma -\omega_1)}{\omega_2^2 \sigmabar^5}\Big\{6m_B \sigmabar \sigma [m_B^2 \sigmabar^2+m_c^2-q^2](m_B \sigma-\omega_1)\nn \\
&+&\omega_ 2\Big[-4m_B \sigma \sigmabar(m_B^2\sigmabar^2-q^2)-2m_B m_c^2(1+\sigma) \sigmabar \nn \\
&+&m_c[(q^2-m_c^2)(1+5\sigma)+m_B^2\sigmabar^2(5\sigma-1)]\Big]\Big\}
{\bar {\bar W}}\nn \\
&+&\frac{8m_B \sigma}{\omega_2^2 \sigmabar^4}(m_B \sigma-\omega_1)\big[  - \omega_2 (m_B^2 \sigmabar^2-q^2) - 8 m_B m_c \omega_2 \sigmabar 
+  2 m_c^2 (3 \omega_1 + \omega_2 - 3 m_B \sigma)  \big]{\bar {\bar Z}}  \nn \\
\eea
\bea
{\bar{ \cal P}}^{3pt(A_1)}_4(\sigma,\omega_1,\omega_2)&=&\frac{12m_c \sigma }{\omega_2 \sigmabar^6}(\omega_1-m_B \sigma)\big[(m_B \sigmabar+m_c)^2-q^2]\big[(m_B \sigmabar-m_c)^2-q^2]{\bar {\bar W}}
\nn \\
&+&\frac{24 m_B m_c^2 \sigma }{\omega_2^2 \sigmabar^5}(m_B \sigma-\omega_1)\Big\{ \big[(m_B \sigmabar-m_c)^2-q^2\big]\omega_2 \nn \\
&+&2(\omega_1-m_B \sigma)[m_B^2 \sigmabar^2+m_c^2-q^2]\Big\}{\bar {\bar Z}} 
\eea
\bea
{\bar{ \cal R}}^{3pt(A_1)}_1(\sigma,\omega_1,\omega_2)&=&\frac{2(m_B \sigma-\omega_1)}{\omega_2 \sigmabar^3}\Big\{\sigmabar \Big[m_c \psi_A-(m_B \sigma +m_c) \psi_V \Big]+(1-3 \sigma){\bar X}_A \Big\} \nn \\
\eea
\bea
&&{\bar{ \cal R}}^{3pt(A_1)}_2(\sigma,\omega_1,\omega_2)= \nn \\
&&\frac{2(m_B \sigma-\omega_1)}{\omega_2^2 \sigmabar^4} \Big\{ \sigmabar \omega_2\big[(m_B \sigmabar+m_c)^2-q^2 \big] \Big[m_c \psi_A-(m_B \sigma+m_c) \psi_V \Big] \nn \\
&-&\omega_2 \Big[m_B^2 \sigmabar^2(3 \sigma-1)+2m_B m_c \sigmabar (1+\sigma)+m_c^2(3+\sigma)+q^2(1-5\sigma) \Big]{\bar X}_A\nn \\
&+&4 m_B \omega_2 \sigma \sigmabar (m_B \sigmabar+m_c)\Big({\bar W}+{\bar Y}_A \Big) \nn \\
&+&4(m_B \sigma-\omega_1) \Big[ \big[m_B(-3 \sigma^2+4 \sigma-1)+m_c(1+2 \sigma)\big]{\bar {\bar W}}-2m_B \sigma \sigmabar {\bar {\bar Z}}  \Big]\Big\} \qquad  \qquad
\eea
\bea
&&{\bar{ \cal R}}^{3pt(A_1)}_3(\sigma,\omega_1,\omega_2)=\nn \\
&&-\frac{4(m_B \sigma-\omega_1)}{\omega_2^2 \sigmabar^5}  \Bigg\{\omega_2 \big[(m_B \sigmabar+m_c)^2-q^2\big] \Bigg[\Big(\sigma \big(m_B^2 \sigmabar^2-q^2-m_c^2)+2m_c^2\Big){\bar X}_A \nn \\
&&-2m_B m_c \sigma \sigmabar\Big({\bar W}+{\bar Y}_A \Big) \Bigg]\nn \\
&&+2 (m_B \sigma-\omega_1)\Bigg[-m_B \sigmabar \Big[m_B \sigmabar (2m_B \sigma \sigmabar+3m_c \sigma +m_c)+2m_c^2(3+\sigma)-2q^2 \sigma \Big]\nn \\
&&+m_c [m_c^2(\sigma-7) +q^2(1+5\sigma)] \Bigg]
 {\bar {\bar W}} +4m_B (m_B \sigma-\omega_1)\sigma \sigmabar  \big( m_B^2 \sigmabar^2-q^2-2m_c^2 \big){\bar {\bar Z}} \Bigg\} \nn \\
\eea
\bea
{\bar{ \cal R}}^{3pt(A_1)}_4(\sigma,\omega_1,\omega_2)&=&\frac{24m_c(m_B \sigma-\omega_1)^2}{\omega_2^2 \sigmabar^6}\big[(m_B \sigmabar+m_c)^2-q^2\big]\nn \\
&\times& \Big\{ \big[\sigma(m_B^2 \sigmabar^2-q^2)+m_c^2(2-\sigma) \big] {\bar {\bar W}}+2m_B m_c \sigma \sigmabar {\bar {\bar Z}} \Big\} \qquad 
\eea
\vskip 0.5cm
\item{$k_{A_2}$}
\bea
c^{had}_{A_2} &=& -i\,f_{H^*_2} m_{H^*_2}^2 \sqrt{m_B m_{H^*_2}}  \,\, , \quad
c^{OPE}_{A_2} =8i \, f_B m_B^3 \,\, , \quad
c_{A_2}=-\frac{8 f_B m_B^{5/2}}{f_{H^*_2} m_{H^*_2}^{5/2}} \,\, , \qquad
\eea
\bea
{\bar{ \cal F}}^{2pt(A_2)}_1 (\sigma)&=&-m_B\frac{\sigma^2}{\bar \sigma}\phi_+  \\
{\bar{ \cal F}}^{2pt(A_2)}_2 (\sigma)&=&m_B\frac{\sigma^2}{{\bar \sigma}^2}\big[-8g_++m_B \sigma {\bar \phi}_\pm \big] \\
{\bar{ \cal F}}^{2pt(A_2)}_3 (\sigma)&=&8\frac{m_B \sigma^2}{{\bar \sigma}^3}\big[2m_B \sigma{\bar G}_\pm + m_c^2  g_+ \big]  \\
{\bar{ \cal F}}^{2pt(A_2)}_4 (\sigma)&=&-24 m_c^2 m_B^2\frac{\sigma^3}{{\bar \sigma}^4}\, {\bar G}_\pm  
\eea
\bea
{\bar{ \cal P}}^{3pt(A_2)}_1 (\sigma,\omega_1,\omega_2)&=&0  \\
{\bar{ \cal P}}^{3pt(A_2)}_2(\sigma,\omega_1,\omega_2)&=&\frac{\sigma}{\sigmabar^2}\Big\{(m_B \sigma +2 m_c) \psi_V\nn \\
&-&\frac{1}{\omega_2}\big[2m_c \omega_2+m_B \sigma(2m_B \sigma -2 \omega_1 -\omega_2) \big]\psi_A \Big\}\nn \\
&-&\frac{2 \sigma^2}{\sigmabar^3}\Big[{\tilde {\bar X}}_A+\frac{1}{\omega_2}(2m_B \sigma -2 \omega_1 -\omega_2) {\bar X}_A \Big]
\eea
\bea
&&{\bar{ \cal P}}^{3pt(A_2)}_3(\sigma,\omega_1,\omega_2)=\nn \\
&-&\frac{2 \sigma}{\omega_2 \sigmabar^4}\big[m_c^2 (2 -\sigma ) + (m_B^2 \sigmabar^2-q^2 ) \sigma \big]\Big[(2m_B \sigma-2\omega_1-\omega_2){\bar X}_A+{\tilde {\bar X}}_A\Big]\nn \\
&-&\frac{4m_B \sigma^2}{\omega_2 \sigmabar^3}\Big\{\big[2 m_c \omega_2 + m_B \sigma ( 2 m_B \sigma -2 \omega_1 - \omega_2 ) \big]\big({\bar Y}_A+{\bar W} \big)+\omega_2(m_B \sigma+2m_c){\tilde {\bar Y}}_A\Big\}\nn \\
&-&\frac{4 \sigma}{\omega_2^2 \sigmabar^4}(m_B \sigma -\omega_1)\big[2 m_c \omega_2 (2 + \sigma) + m_B  \sigma (1 + 2  \sigma) ( 2 m_B \sigma -2 \omega_1 - \omega_2 )\big]{\bar {\bar W}}\nn \\
&-& \frac{8m_B\sigma^2}{\omega_2^2\sigmabar^3}(m_B \sigma -\omega_1)(2m_B \sigma-2\omega_1-\omega_2){\bar{\bar Z}}
\eea
\bea
&&{\bar{ \cal P}}^{3pt(A_2)}_4(\sigma,\omega_1,\omega_2)=\nn \\
&&\frac{24m_B m_c \sigma^2}{\omega_2^2\sigmabar^4}(m_B \sigma -\omega_1)\big[m_c(2m_B \sigma-2\omega_1-\omega_2)+\omega_2 2 m_B \sigma \big]{\bar{\bar Z}}\nn \\
&+&\frac{12 \sigma (m_B \sigma -\omega_1)}{\omega_2^2 \sigmabar^5}\Big\{-2m_B \sigma(m_B \sigma -\omega_1)\big[m_c^2 (2 -\sigma) + (m_B^2 \sigmabar^2-q^2)\sigma \big]\nn \\
&+&\omega_2\big[m_B \sigma^2(m_B^2 \sigmabar^2-q^2)+m_c[-2\sigma(m_B^2 \sigmabar-q^2)-2m_c^2+m_B m_c \sigma (2-\sigma)]\big] \Big\}{\bar {\bar W}} \qquad \nn \\
\eea
\bea
{\bar{ \cal R}}^{3pt(A_2)}_1 (\sigma,\omega_1,\omega_2)&=&0 \\
{\bar{ \cal R}}^{3pt(A_2)}_2(\sigma,\omega_1,\omega_2)&=&-\frac{2\sigma^2(m_B \sigma-\omega_1)}{\omega_2 \sigmabar^3}\Big\{m_B \sigmabar \psi_A +2 {\bar X}_A \Big\}
\eea
\bea
{\bar{ \cal R}}^{3pt(A_2)}_3(\sigma,\omega_1,\omega_2)&=&
\frac{4\sigma(m_B \sigma-\omega_1)}{\omega_2^2 \sigmabar^4}
\Big\{-2m_B^2 \omega_2 \sigmabar \sigma^2 
 \big[{\bar Y}_A+{\bar W} \big] \nn \\
 &-&\omega_2 \Big[\sigma \big( m_B^2 \sigmabar^2-q^2-m_c^2 \big) +2m_c^2 \Big]{\bar X}_A \nn \\
 &-&2m_B \sigma (m_B \sigma-\omega_1) \Big[(1+2 \sigma){\bar {\bar W}}+2\sigmabar {\bar {\bar Z}}\Big] \Big\}
 \eea
 \bea
 {\bar{ \cal R}}^{3pt(A_2)}_4(\sigma,\omega_1,\omega_2)&=&
 -\frac{24m_B \sigma^2(m_B \sigma-\omega_1)^2}{\omega_2^2 \sigmabar^5} \nn \\
 &\times&
\Big\{  \Big[\sigma \big( m_B^2 \sigmabar^2-q^2-m_c^2 \big) +2m_c^2 \Big]{\bar {\bar W}}-2m_c^2\sigmabar {\bar {\bar Z}} \Big\} \qquad 
\eea
\vskip 0.5cm
\item{$k_{A_3}$}
\bea
c^{had}_{A_3} &=& -i\, f_{H^*_2} m_{H^*_2}^2 \sqrt{m_B m_{H^*_2}}  \,\, , \quad
c^{OPE}_{A_3} =4i \,f_B m_B^2  \,\, , \quad
c_{A_3}=-\frac{4 f_B m_B^{3/2}}{f_{H^*_2} m_{H^*_2}^{5/2}}  \,\, , \qquad
\eea
\bea
{\bar{ \cal F}}^{2pt(A_3)}_1 (\sigma)&=&m_Bm_{H^*_2}\frac{\sigma}{\bar \sigma}\phi_+  \\
{\bar{ \cal F}}^{2pt(A_3)}_2 (\sigma)&=&m_Bm_{H^*_2}\frac{\sigma}{{\bar \sigma}^2}\big[8g_+ +(m_c-2m_B \sigma) {\bar \phi}_\pm\big] \\
{\bar{ \cal F}}^{2pt(A_3)}_3 (\sigma)&=&-8m_Bm_{H^*_2} \frac{\sigma}{{\bar \sigma}^3} \Big[(4m_B \sigma-m_c) {\bar G}_\pm +m_c^2g_+ \Big] \\
{\bar{ \cal F}}^{2pt(A_3)}_4 (\sigma)&=&24 m_B m_{H^*_2}  m_c^2\frac{\sigma}{\sigmabar^4}(2m_B \sigma-m_c) {\bar G}_\pm 
\eea
\bea
{\bar{ \cal P}}^{3pt(A_3)}_1 (\sigma,\omega_1,\omega_2)&=&0  \\
{\bar{ \cal P}}^{3pt(A_3)}_2(\sigma,\omega_1,\omega_2)&=&m_B m_{H^*_2}\frac{\sigma}{\omega_2\sigmabar^2}\nn \\
&\times&\Big[(2m_B \sigma -2\omega_1+\omega_2)\psi_A-(4m_B \sigma-4\omega_1-\omega_2)\psi_V \Big]\nn \\
&- &m_{H^*_2}\frac{1}{\omega_2\sigmabar^3}\Big\{\big[\omega_2(1+\sigma)-4\sigma(m_B \sigma-\omega_1)\big]{\bar X}_A\nn \\
&+& \big[2m_B\sigma \sigmabar-2\omega_1\sigmabar-\omega_2(1+\sigma)\big]
{\tilde {\bar X}}_A\Big\}
\eea
\bea
{\bar{ \cal P}}^{3pt(A_3)}_3(\sigma,\omega_1,\omega_2)&=&
\frac{2m_{H^*_2} \sigma}{\omega_2 \sigmabar^4}\Big\{ \Big[2(m_B^2 \sigmabar^2+m_c^2-q^2)(m_B \sigma-\omega_1 )\nn \\
&-&\omega_2 \big[(m_B \sigmabar-m_c)^2-q^2]\Big]{\bar X}_A\nn \\
&+&\omega_2[(m_B \sigmabar+m_c)^2-q^2]{\tilde {\bar X}}_A \Big\}\nn \\
&+&\frac{4m_B m_{H^*_2}\sigma}{\omega_2 \sigmabar^3}\Big\{\big[ m_c \omega_2 + 2m_B \sigma ( 2 m_B \sigma -2 \omega_1 - \omega_2 ) \big]\big({\bar Y}_A+{\bar W} \big)\nn \\
&+&\omega_2(2m_B \sigma+m_c){\tilde {\bar Y}}_A\Big\}\nn \\
&+&8m_B m_{H^*_2}\frac{\sigma}{\omega_2\sigmabar^3}(m_B \sigma-\omega_1 ){\bar {\bar Z}}\nn \\
&+&4 m_{H^*_2}\frac{1}{\omega_2^2 \sigmabar^4}(m_B \sigma-\omega_1 )\big[2m_B \sigma (1+5 \sigma) (m_B \sigma-\omega_1)\nn \\
&-&\omega_2(1+2 \sigma)(2m_B \sigma-m_c) \big] {\bar {\bar W}}
\eea
\bea
{\bar{ \cal P}}^{3pt(A_3)}_4(\sigma,\omega_1,\omega_2)&=&
24m_Bm_{H^*_2} m_c\, \frac{\sigma}{\omega_2^2\sigmabar^4}(m_B \sigma -\omega_1)\nn \\
&\times&\big[m_c(2m_B \sigma-2\omega_1-\omega_2)-4m_B\sigma  \omega_2\big]{\bar{\bar Z}}\nn \\
&+&12m_{H^*_2} \, \frac{\sigma}{\omega_2^2\sigmabar^5}(m_B \sigma -\omega_1)\Big[4m_B [\sigma(m_B^2 \sigmabar^2-q^2)+m_c^2](m_B \sigma -\omega_1)\nn \\
&-&(2m_B \sigma -m_c)\big[ (m_B \sigmabar-m_c)^2-q^2 \big]\omega_2 \Big]{\bar {\bar W}} 
\eea
\bea
{\bar{ \cal R}}^{3pt(A_3)}_1 (\sigma,\omega_1,\omega_2)&=&0 \\
{\bar{ \cal R}}^{3pt(A_3)}_2(\sigma,\omega_1,\omega_2)&=&m_{H^*_2}\frac{2(m_B\sigma-\omega_1)}{\omega_2\sigmabar^3}\nn \\
&\times&\Big\{\sigmabar \Big[(2m_B \sigma -m_c)\psi_A-(m_B \sigma-m_c) \psi_V\Big]+(3 \sigma -1){\bar X}_A \Big\} \qquad  \nn \\
\eea
\bea
{\bar{ \cal R}}^{3pt(A_3)}_3(\sigma,\omega_1,\omega_2)&=&-m_{H^*_2}\frac{4(m_B\sigma-\omega_1)}{\omega_2^2\sigmabar^4}\Big\{
-\omega_2\Big[\sigma \big( m_B^2 \sigmabar^2-q^2-m_c^2 \big) +2m_c^2 \Big]{\bar X}_A\nn \\
&-&2m_B \omega_2 \sigma \sigmabar(2m_B \sigma-m_c)\big({\bar Y}_A+{\bar W} \big) \nn \\
&+&2(m_B \sigma-\omega_1) \Big[\Big(m_c(1+2 \sigma)-6m_B \sigma^2 \Big){\bar {\bar W}}-2m_B \sigma \sigmabar{\bar {\bar Z}} \Big] \Big\} \qquad \nn \\
\eea
\bea
{\bar{ \cal R}}^{3pt(A_3)}_4(\sigma,\omega_1,\omega_2)&=&m_{H^*_2}\frac{24(m_B\sigma-\omega_1)^2}{\omega_2^2\sigmabar^5}\Big\{-2m_B m_c^2\sigma \sigmabar 
{\bar {\bar Z}}\nn \\
 &+&(2m_B \sigma -m_c) \Big[\sigma \big( m_B^2 \sigmabar^2-q^2-m_c^2 \big) +2m_c^2 \Big]{\bar {\bar W}} \Big\} 
\eea
\vskip 0.5cm
\item{$k_P$}
\bea
c^{had}_P &=&  f_{H^*_2} m_{H^*_2}^2 \sqrt{m_B m_{H^*_2}}  \,\, , \quad
c^{OPE}_P =4 \,f_B m_B^2  \,\, , \quad
c_P=\frac{4 f_B m_B^{3/2}}{f_{H^*_2} m_{H^*_2}^{5/2}} \,\, , \qquad
\eea
\bea
{\bar{ \cal F}}^{2pt(P)}_1 (\sigma)&=&-m_B\frac{\sigma}{\bar \sigma}(m_B \sigma+m_c)\phi_+ -m_B\frac{\sigma^2}{ \sigmabar^2} {\bar \phi}_\pm  \\
{\bar{ \cal F}}^{2pt(P)}_2 (\sigma)&=&-4m_B\frac{\sigma}{\sigmabar^2}(2m_B \sigma+m_c)g_+ 
-8m_B\frac{\sigma (1+\sigma)}{ \sigmabar^3}{\bar G}_\pm \nn \\
&-&m_B\frac{\sigma}{ \sigmabar^3}\big[(m_B^2 \sigmabar^2-q^2)\sigma-m_B m_c \sigma \sigmabar+m_c^2 \big]{\bar \phi}_\pm \\
{\bar{ \cal F}}^{2pt(P)}_3 (\sigma)&=&8m_B m_c^2\frac{\sigma}{\sigmabar^3}(m_B \sigma+m_c)g_+ \nn \\
&+&8m_B\frac{\sigma^2}{ \sigmabar^4}\big[-2m_B^2 \sigmabar^2+m_B m_c \sigmabar +m_c^2+2q^2 \big]{\bar G}_\pm  \\
{\bar{ \cal F}}^{2pt(P)}_4 (\sigma)&=&24m_B m_c^2\frac{\sigma}{ \sigmabar^5}\big[(m_B^2 \sigmabar^2-q^2)\sigma-m_B m_c \sigma \sigmabar+m_c^2 \big]{\bar G}_\pm 
\eea
\bea
{\bar{ \cal P}}^{3pt(P)}_1 (\sigma,\omega_1,\omega_2)&=&\frac{2 \sigma (m_B\sigma-\omega_1)}{\omega_2\sigmabar^2}\big[-\psi_V+\psi_A \big]  
\\ 
{\bar{ \cal P}}^{3pt(P)}_2(\sigma,\omega_1,\omega_2)&=&\frac{ \sigma}{\omega_2 \sigmabar^3}\Big\{\Big[2(m_B \sigma-\omega_1)[m_B^2 \sigmabar(1-2\sigma)+m_c^2-q^2]\nn \\
&+&3m_B \omega_2 \sigmabar(m_B \sigma-m_c)\Big]\psi_A \nn \\
&-&\Big[2(m_B \sigma-\omega_1)[m_B^2 (1+\sigma) \sigmabar+m_c^2-q^2] \nn \\
&-&3m_B \omega_2\sigmabar(m_B \sigma+m_c)\Big]\psi_V \Big\} \nn \\
&-&\frac{1}{\omega_2 \sigmabar^3}\Big\{ \big[-m_c \omega_2 (1 + \sigma) + 2 m_B \sigma^2 (2 m_B \sigma-2 \omega_1 - \omega_2 )\big] {\bar X}_A \nn \\
&+&\omega_2[2m_B \sigma^2-m_c(1+\sigma)]{\tilde {\bar X}}_A \Big\} \nn \\
&+&\frac{2m_B \sigma}{\omega_2 \sigmabar^3}\Big\{\big[4m_B \sigma-4 \omega_1+\omega_2(-3+\sigma)\big]\big({\bar Y}_A+{\bar W} \big)\nn \\
&-&\big[2 \sigmabar(m_B \sigma-\omega_1)+\omega_2(-3+\sigma) \big]{\tilde {\bar Y}}_A\Big\} \nn \\
&+&\frac{2 }{\omega_2^2 \sigmabar^4}(m_B \sigma -\omega_1)\big[4\sigma(2+\sigma)(m_B \sigma-\omega_1)-\omega_2(1+4\sigma+\sigma^2)\big]{\bar {\bar W}} \nn \\
\eea
\bea
{\bar{ \cal P}}^{3pt(P)}_3 (\sigma,\omega_1,\omega_2)&=&\frac{2 \sigma }{\omega_2\bar{\sigma}^4}\Big\{-2m_B(m_B \sigma-\omega_1) \Big[\sigma (m_B^2 \sigmabar^2-q^2)-m_c^2(\sigma-2) \Big]\nn \\
&+&\omega_2(m_B \sigma+m_c)\Big[(m_B \sigmabar+m_c)^2-q^2\Big] \Big\}{\bar X}_A\nn\\
&-&\frac{2 \sigma }{\bar{\sigma}^4}(m_B \sigma-m_c)[(m_B \sigmabar-m_c)^2-q^2]{\tilde {\bar X}}_A\nn \\
&+&\frac{4m_B \sigma }{\omega_2\bar{\sigma}^4}\Big\{2(m_B \sigma-\omega_1) \Big[\sigma (m_B^2 \sigmabar^2-q^2)-m_c^2(1-2\sigma) \Big]\nn \\
&-&\omega_2\Big[\sigma[(m_B \sigmabar+m_c)^2-q^2]+m_c \sigmabar(m_B \sigma-m_c)\Big] \Big\}[{\bar W}+{\bar Y}_A]\nn \\
&+&\frac{4m_B \sigma }{\bar{\sigma}^4}\Big\{\sigma[(m_B \sigmabar-m_c)^2-q^2]-m_c \sigmabar(m_B \sigma+m_c)\Big\}{\tilde {\bar Y}}_A\nn\\
&-&\frac{8m_B \sigma }{\omega_2^2\bar{\sigma}^4}(m_B \sigma-\omega_1)\Big\{4m_B(m_B \sigma-\omega_1)\sigma \sigmabar\nn \\
&+&\omega_2[2m_c(2+\sigma)-3m_B \sigma \sigmabar] \Big\} {\bar {\bar Z}}\nn\\
&-&\frac{4(m_B \sigma -\omega_1)}{\omega_2^2\bar{\sigma}^5}\Big\{-2\sigma(m_B \sigma-\omega_1)\big[m_B^2 \sigmabar^2(2+3 \sigma)\nn \\
&-&2q^2(1+2\sigma)+m_c^2(5+\sigma)\big]\nn\\
&+&\omega_2 \Big[3\sigma(1+\sigma)(m_B^2\sigmabar^2-q^2)-m_B m_c \sigma(\sigma^2+\sigma-2)\nn \\
&+&m_c^2(\sigma^2+4\sigma+1) \Big] \Big\} {\bar {\bar W}}
\eea
\bea
{\bar{ \cal P}}^{3pt(P)}_4(\sigma,\omega_1,\omega_2)&=&-\frac{12  \sigma}{\omega_2^2 \sigmabar^6}(m_B \sigma-\omega_1)\Big\{(m_B \sigma -\omega_1) \Big[-2\sigma (m_B^2\sigmabar^2-q^2)^2 \nn  \\
&-&2m_c^2\big[m_B^2 \sigmabar^3-q^2(1+\sigma)+m_c^2 \big]\Big]\nn \\
&+&\omega_2\big[(m_B \sigmabar+m_c)^2-q^2 \big] \big[\sigma(m_B^2 \sigmabar^2-q^2)-m_B m_c \sigma \sigmabar+m_c^2 \big] \Big\}{\bar {\bar W}}\nn \\
&-&\frac{24 m_B m_c \sigma}{\omega_2^2 \sigmabar^5}(m_B \sigma-\omega_1)\Big\{-6m_B m_c \sigma \sigmabar (m_B \sigma-\omega_1)\nn \\ 
&+&\omega_2 \Big[2 \sigma (m_B^2 \sigmabar^2-q^2)+3m_B m_c  \sigma \sigmabar+m_c^2 (1+\sigma)\Big] \Big\} {\bar {\bar Z}}
\eea
\bea
{\bar{ \cal R}}^{3pt(P)}_1(\sigma,\omega_1,\omega_2)&=&\frac{2  \sigma (m_B \sigma-\omega_1)}{\omega_2 \sigmabar^2}\big( \psi_A -\psi_V \big) \\
{\bar{ \cal R}}^{3pt(P)}_2(\sigma,\omega_1,\omega_2)&=&-\frac{2   (m_B \sigma-\omega_1)}{\omega_2^2 \sigmabar^4}\nn \\
&\times&\Big\{\omega_2 \sigmabar \Big[(m_B^2 \sigma \sigmabar-q^2 \sigma+m_c^2) \psi_V \nn \\
&+&\big[m_B \sigma \sigmabar (m_c-m_B \sigmabar)+q^2 \sigma-m_c^2 \big] \psi_A \Big] \nn \\
&+&\omega_2 \sigmabar \big[2m_B \sigma^2+m_c(1+\sigma) \big]{\bar X}_A-2m_B \omega_2 \sigma \sigmabar(1+\sigma)\big({\bar W}+{\bar Y}_A \big)\nn \\
&+&2 (m_B \sigma-\omega_1) \big[1-\sigma(4+3\sigma) \big]{\bar {\bar W}} \Big\}
\eea
\bea
{\bar{ \cal R}}^{3pt(P)}_3(\sigma,\omega_1,\omega_2)&=&-\frac{4   (m_B \sigma-\omega_1)}{\omega_2^2 \sigmabar^5}\nn \\
&\times& \Big\{\omega_2(m_B \sigma+m_c) \sigmabar \big[m_B^2 \sigma \sigmabar^2-q^2 \sigma +m_c^2(2-\sigma) \big]{\bar X}_A\nn \\
&-&2m_B \omega_2 \sigma \sigmabar \big[m_B \sigma \sigmabar(m_B \sigmabar-m_c)-q^2 \sigma+m_c^2 \big]\big({\bar W}+{\bar Y}_A \big)\nn \\
&-&2 (m_B \sigma-\omega_1)\big[ m_B^2 \sigma \sigmabar^2(1+3\sigma)-m_B m_c \sigma \sigmabar(2+\sigma)-q^2\sigma(1+5\sigma)\nn \\
&+&m_c^2(3+4\sigma-\sigma^2)\big]{\bar {\bar W}}\nn \\
&+&4m_B^2 \sigma^2 \sigmabar^2(m_B \sigma-\omega_1){\bar {\bar Z}} \Big\} 
\eea
\bea
{\bar{ \cal R}}^{3pt(P)}_4(\sigma,\omega_1,\omega_2)&=&\frac{24  (m_B \sigma-\omega_1)^2}{\omega_2^2 \sigmabar^6}\Big\{ 2m_B m_c^2 \sigma \sigmabar^2(m_B \sigma +m_c){\bar {\bar Z}} \nn \\
&+& \big[\sigma (m_B^2\sigmabar^2-q^2-m_c^2)+2m_c^2 \big]\big[m_B \sigma \sigmabar(m_B \sigmabar-m_c)-q^2 \sigma +m_c^2 \big]{\bar {\bar W}}\Big\} \nn \\
\eea
\vskip 0.5cm
%
\item{ $k_{T_1}$}
\bea
c^{had}_{T_1} &=& \frac{1}{2} f_{H^*_2} m_{H^*_2}^2 \sqrt{m_B m_{H^*_2}}  \,\, , \quad
c^{OPE}_{T_1} =2 \,f_B m_B^2 \,\, , \quad
c_{T_1}=\frac{4 f_B m_B^{3/2}}{f_{H^*_2} m_{H^*_2}^{5/2}} \,\, , \qquad
\eea

\bea
{\bar{ \cal F}}^{2pt(T_1)}_1 (\sigma)&=&m_B\frac{\sigma}{\bar \sigma}(m_B \sigma-m_c)\phi_+   \\
{\bar{ \cal F}}^{2pt(T_1)}_2 (\sigma)&=&m_B\frac{\sigma}{\sigmabar^2}\big[4(2m_B \sigma-m_c)g_+ 
-8{\bar G}_\pm +m_c(m_B \sigma -m_c) {\bar \phi}_\pm \big] \\
{\bar{ \cal F}}^{2pt(T_1)}_3 (\sigma)&=&-8m_B m_c\frac{\sigma}{\sigmabar^3}\big[m_c(m_B \sigma-m_c)g_+ 
-m_B \sigma{\bar G}_\pm \big] \\
{\bar{ \cal F}}^{2pt(T_1)}_4 (\sigma)&=&-24m_B m_c^3\frac{\sigma}{ \sigmabar^4}(m_B \sigma-m_c){\bar G}_\pm
\eea

\bea
{\bar{ \cal P}}^{3pt(T_1)}_1 (\sigma,\omega_1,\omega_2)&=&0 
\eea
\bea
{\bar{ \cal P}}^{3pt(T_1)}_2(\sigma,\omega_1,\omega_2)&=&m_B \frac{\sigma}{\omega_2 \sigmabar^2}\nn \\
&\times&\Big\{ \omega_2(m_B \sigma -m_c) \psi_V+\big[m_B \sigma(2\omega_1+\omega_2-2m_B \sigma)+m_c \omega_2\big]\psi_A \nn \\
&+&2(2\omega_1+3\omega_2-2m_B \sigma){\tilde {\bar Y}}_A-2(4\omega_1+3\omega_2-4m_B \sigma)[{\bar W}+{\bar Y}_A] \Big\}\nn \\
&-&\frac{1}{\omega_2^2 \sigmabar^3}\Big\{ \omega_2[2m_B\sigma^2(2\omega_1+\omega_2-2m_B \sigma)+m_c \omega_2(3 \sigma-1)]{\bar X}_A
\nn \\
&-&\omega_2^2[2m_B \sigma^2-m_c(3\sigma-1)]{\tilde {\bar X}}_A \nn \\
&+&2(m_B \sigma -\omega_1)[\omega_2(1+5\sigma)-8\sigma(m_B \sigma -\omega_1)]{\bar {\bar W}} \Big\}
\eea
\bea
{\bar{ \cal P}}^{3pt(T_1)}_3(\sigma,\omega_1,\omega_2)&=&\frac{2 \sigma}{\omega_2 \sigmabar^4 }\Big\{ 
2m_B(m_B \sigma-\omega_1)\big[\sigma(m_B^2\sigmabar^2-m_c^2-q^2)+2m_c^2 \big]
\nn \\
&-&\omega_2(m_B \sigma+m_c)\big[(m_B \sigmabar+m_c)^2-q^2 \big]\Big\}{\bar X}_A
\nn \\
&+&2\frac{\sigma}{\sigmabar^4}(m_B \sigma-m_c)[(m_B \sigmabar-m_c)^2-q^2]{\tilde {\bar X}}_A  \nn \\
&+&4m_B \frac{\sigma}{\omega_2 \sigmabar^3}
\Big\{ \omega_2 m_c(m_B \sigma-m_c){\tilde {\bar Y}}_A
\nn \\
&+&m_c \big[m_c(2\omega_1+\omega_2-2m_B \sigma)+\omega_2 m_B \sigma \big]\big({\bar W}+{\bar Y}_A \big)
\nn \\ 
&+&2(m_B \sigma -4m_c)(m_B \sigma-\omega_1) {\bar {\bar Z}} \Big\}
\nn \\
&-&\frac{4(m_B \sigma-\omega_1) }{\omega_2^2 \sigmabar^4}\Big\{2\sigma (m_B \sigma-\omega_1)\big[m_B^2 \sigmabar(\sigma-2)+m_c^2+2q^2\big]
\nn \\
&+&\omega_2 \big[m_c^2 \sigmabar - 3 q^2 \sigma + 3 m_B m_c (-2 + \sigma) \sigma + 
 m_B^2 \sigma  \sigmabar (3 - \sigma)\big]\Big\}{\bar {\bar W}} \qquad  \qquad 
\eea
\newpage
\bea
{\bar{ \cal P}}^{3pt(T_1)}_4(\sigma,\omega_1,\omega_2)&=&-\frac{12  m_c \sigma}{\omega_2^2 \sigmabar^5}(m_B \sigma -\omega_1)\Big[2m_c(m_B \sigma -\omega_1)\big[m_B^2 \sigmabar(1+\sigma)+m_c^2-q^2 \big]\nn \\
&-&\omega_2(m_B \sigma+m_c)\big[(m_B \sigmabar+m_c)^2-q^2 \big]\Big]{\bar {\bar W}}
\nn \\
&+&\frac{24 m_B m_c^2 \sigma}{\omega_2^2 \sigmabar^4}(m_B \sigma -\omega_1)\big[m_B \sigma (2m_B \sigma-2\omega_1 -\omega_2)-m_c \omega_2 \big]{\bar {\bar Z}}  \nn \\
\eea
\bea
{\bar{ \cal R}}^{3pt(T_1)}_1 (\sigma,\omega_1,\omega_2)&=&0 \eea
\bea
{\bar{ \cal R}}^{3pt(T_1)}_2 (\sigma,\omega_1,\omega_2)&=&-\frac{2(m_B \sigma -\omega_1)}{\omega_2^2 \sigmabar^3}\nn \\
&\times& \Big\{\omega_2\sigmabar(m_B \sigma -m_c)\big[m_c \psi_A-(m_c+m_B \sigma)\psi_V \big] 
\nn \\
&+&\omega_2 \big[ m_c(1+\sigma)-2m_B \sigma^2 \big]{\bar X}_A-2m_B \omega_2 \sigma \sigmabar \big({\bar W}+{\bar Y}_A \big) \nn \\
&-&2(m_B \sigma -\omega_1) (3 \sigma-1){\bar {\bar W}} \Big\} 
\eea
\bea
{\bar{ \cal R}}^{3pt(T_1)}_3 (\sigma,\omega_1,\omega_2)&=&-\frac{4(m_B \sigma -\omega_1)}{\omega_2^2 \sigmabar^4}\nn \\
&\times&\Big\{-\omega_2 (m_B \sigma-m_c) \big[\sigma (m_B^2 \sigmabar^2-q^2-m_c^2)+2m_c^2 \big]{\bar X}_A \nn \\
&+&2m_B m_c \omega_2 \sigma \sigmabar (m_B \sigma-m_c)\big({\bar W}+{\bar Y}_A \big)-4m_B^2 \sigma^2 \sigmabar (m_B \sigma -\omega_1) {\bar {\bar Z}} \nn \\
&-&2(m_B \sigma-\omega_1) \big[m_B^2 \sigma \sigmabar^2-m_B m_c \sigma (2+\sigma)+m_c^2(3+\sigma)-q^2 \sigma \big]{\bar {\bar W}} \Big\} \nn \\
\eea
\bea
{\bar{ \cal R}}^{3pt(T_1)}_4 (\sigma,\omega_1,\omega_2)&=&-\frac{24  m_c (m_B \sigma -\omega_1)^2}{\omega_2^2 \sigmabar^5}(m_B \sigma -m_c) \nn \\
&\times&\Big\{\big[\sigma(m_B^2 \sigmabar^2-q^2-m_c^2)+2m_c^2 \big]{\bar {\bar W}} +2m_B m_c \sigma \sigmabar {\bar {\bar Z}} \Big\} \qquad
\eea
\vskip 0.5cm
%
\item{ $k_{T_2}$}
\bea
c^{had}_{T_2} &=&   \frac{1}{2} f_{H^*_2} m_{H^*_2}^2 \sqrt{m_B m_{H^*_2} } \,\, , \quad
c^{OPE}_{T_2} =2 \,f_B m_B^2 m_{H^*_2} \,\, , \quad
c_{T_2}=\frac{4 f_B m_B^{3/2}}{f_{H^*_2} m_{H^*_2}^{3/2}}  \,\, , \qquad
\eea
\bea
{\bar{ \cal F}}^{2pt(T_2)}_1 (\sigma)&=&-m_B\frac{\sigma}{\bar \sigma}\phi_+   \\
{\bar{ \cal F}}^{2pt(T_2)}_2 (\sigma)&=&-m_B\frac{\sigma}{\sigmabar^2}\big[8g_+ +m_c {\bar \phi}_\pm \big]\\
{\bar{ \cal F}}^{2pt(T_2)}_3 (\sigma)&=&8m_B m_c\frac{\sigma}{\sigmabar^3}\big[m_c\,g_+  - {\bar G}_\pm \big] \\
{\bar{ \cal F}}^{2pt(T_2)}_4 (\sigma)&=&24m_B m_c^3\frac{\sigma}{ \sigmabar^4}{\bar G}_\pm 
\eea
\bea
{\bar{ \cal P}}^{3pt(T_2)}_1 (\sigma,\omega_1,\omega_2)&=&0  \\
{\bar{ \cal P}}^{3pt(T_2)}_2(\sigma,\omega_1,\omega_2)&=&-m_B \frac{\sigma }{\omega_2 \sigmabar^2}\big[(2 \omega_1+\omega_2-2m_B \sigma)\psi_A+\omega_2 \psi_V \big]
\nn \\
&+&\frac{1}{\omega_2 \sigmabar^3}\Big\{[\omega_2(1+\sigma)+4\sigma(\omega_1-m_B \sigma)]{\bar X}_A \nn \\
&-& [\omega_2(1+\sigma)+2\sigmabar(\omega_1-m_B \sigma)]{\tilde {\bar X}}_A\Big\}  
\eea
\bea
{\bar{ \cal P}}^{3pt(T_2)}_3(\sigma,\omega_1,\omega_2)&=&\frac{2\sigma}{\omega_2 \sigmabar^4}\Big\{\Big[-2(m_B \sigma -\omega_1)(m_B^2\sigmabar^2+m_c^2-q^2)\nn \\
&+&\omega_2\big[(m_B\sigmabar+m_c)^2-q^2\big]\Big]{\bar X}_A
-\omega_2 \big[(m_B\sigmabar+m_c)^2-q^2\big]{\tilde {\bar X}}_A\Big\}\nn \\
&-&\frac{4m_B  \sigma}{\sigmabar^3}\Big\{m_c \Big( {\bar W}+{\bar Y}_A+{\tilde {\bar Y}}_A \big)+2\frac{m_B \sigma -\omega_1}{\omega_2} {\bar {\bar Z}} \Big\}\nn \\
&-&\frac{4(m_B \sigma -\omega_1)}{\omega_2^2 \sigmabar^4} \Big[ 2m_B \sigma \sigmabar(m_B \sigma -\omega_1)\nn \\
&-&\omega_2 [2m_B \sigma \sigmabar-m_c(1+2\sigma)] \Big]{\bar {\bar W}} 
\eea
\bea
{\bar{ \cal P}}^{3pt(T_2)}_4(\sigma,\omega_1,\omega_2)&=&\frac{12 m_c \sigma}{\omega_2^2 \sigmabar^5}(m_B \sigma -\omega_1) \Big[4m_B m_c \sigmabar(m_B \sigma-\omega_1)\nn \\
&-&\omega_2\big[(m_B \sigmabar+m_c)^2-q^2 \big] \Big]{\bar {\bar W}}\nn \\
&+&\frac{24 m_B m_c^2 \sigma}{\omega_2^2 \sigmabar^4}(m_B \sigma -\omega_1) (2\omega_1 +\omega_2 -2m_B \sigma){\bar {\bar Z}} \qquad
\eea
\bea
{\bar{ \cal R}}^{3pt(T_2)}_1 (\sigma,\omega_1,\omega_2)&=&0  \\
{\bar{ \cal R}}^{3pt(T_2)}_2 (\sigma,\omega_1,\omega_2)&=&\frac{2 (m_B \sigma-\omega_1) }{\omega_2 \sigmabar^3} \nn \\
&\times&\Big\{
\sigmabar \big[m_c \psi_A -(m_B \sigma+m_c) \psi_V \big]+(1-3\sigma){\bar X}_A \Big\}  \\
{\bar{ \cal R}}^{3pt(T_2)}_3 (\sigma,\omega_1,\omega_2)&=&-\frac{4 (m_B \sigma-\omega_1) }{\omega_2^2 \sigmabar^4}\Big\{\omega_2 \big[ \sigma(m_B^2\sigmabar^2-q^2-m_c^2)+2m_c^2 \big]{\bar X}_A
\nn \\
&-&2m_B m_c \omega_2 \sigma \sigmabar \big( {\bar W}+{\bar Y}_A \big) \nn \\
&+&2(m_B \sigma-\omega_1)\Big[-m_c(1+2\sigma){\bar {\bar W}}+2m_B \sigma \sigmabar{\bar {\bar Z}} \Big] \Big\} \\
{\bar{ \cal R}}^{3pt(T_2)}_4(\sigma,\omega_1,\omega_2)&=&\frac{24m_c (m_B \sigma-\omega_1)^2 }{\omega_2^2 \sigmabar^5} \nn \\
&\times&\Big\{
\Big[\sigma (m_B^2 \sigmabar^2-q^2)+m_c^2(2-\sigma) \big]{\bar {\bar W}}+2m_B m_c\sigma \sigmabar {\bar {\bar Z}}\Big\} 
\eea
\vskip 0.5cm
%
\item{$k_{T_3}$}
\bea
c^{had}_{T_3} &=&   f_{H^*_2} m_{H^*_2}^2 \sqrt{m_B m_{H^*_2} } \,\, , \quad
c^{OPE}_{T_3} =8 \,f_B m_B^3 m_{H^*_2}  \,\, , \quad
c_{T_3}=\frac{8 f_B m_B^{5/2}}{f_{H^*_2} m_{H^*_2}^{3/2}}  \,\, , \qquad
\eea
\bea
{\bar{ \cal F}}^{2pt(T_3)}_1 (\sigma)&=&0  \\
{\bar{ \cal F}}^{2pt(T_3)}_2 (\sigma)&=&m_B\frac{\sigma^2}{\sigmabar^2}{\bar \phi}_\pm \\
{\bar{ \cal F}}^{2pt(T_3)}_3 (\sigma)&=&16m_B\frac{\sigma^2}{\sigmabar^3}{\bar G}_\pm \\
{\bar{ \cal F}}^{2pt(T_3)}_4 (\sigma)&=&-24m_B m_c^2\frac{\sigma^2}{ \sigmabar^4}{\bar G}_\pm
\eea
\bea
{\bar{ \cal P}}^{3pt(T_3)}_1 (\sigma,\omega_1,\omega_2)&=&0  \\
{\bar{ \cal P}}^{3pt(T_3)}_2(\sigma,\omega_1,\omega_2)&=&2\frac{\sigma}{\omega_2 \sigmabar^2}(m_B \sigma -\omega_1)\big[\psi_V - \psi_A \big]  \\
{\bar{ \cal P}}^{3pt(T_3)}_3(\sigma,\omega_1,\omega_2)&=&-4m_c \frac{\sigma}{\sigmabar^3}\big[{\bar X}_A+{\tilde {\bar X}}_A \big]\nn \\
&+&4m_B \frac{\sigma^2}{\omega_2\sigmabar^3} \Big[(2 \omega_1+\omega_2-2m_B \sigma)\, \big[{\bar Y}_A+{\bar W}\big]
-\omega_2{\tilde {\bar Y}}_A\Big]  \qquad \nn \\
&+&12 \frac{\sigma^2}{\omega_2^2\sigmabar^4} (m_B \sigma-\omega_1)(2 \omega_1+\omega_2-2m_B \sigma){\bar {\bar W}}  \\
{\bar{ \cal P}}^{3pt(T_3)}_4(\sigma,\omega_1,\omega_2)&=&12\frac{\sigma}{\omega_2^2 \sigmabar^5}(m_B \sigma -\omega_1)\nn \\
&\times&\Big[(2 \omega_1+\omega_2-2m_B \sigma)\big[\sigma (m_B^2 \sigmabar^2-m_c^2-q^2)+2m_c^2\big]\nn \\
&-&2\omega_2 m_B m_c \sigma \sigmabar\Big]{\bar {\bar W}}
+48m_B m_c \frac{\sigma^2}{\omega_2 \sigmabar^4}(m_B \sigma-\omega_1) {\bar {\bar Z}}
\eea
\bea
{\bar{ \cal R}}^{3pt(T_3)}_1 (\sigma,\omega_1,\omega_2)&=&0  \\
{\bar{ \cal R}}^{3pt(T_3)}_2(\sigma,\omega_1,\omega_2)&=&-\frac{2\sigma (m_B \sigma-\omega_1) }{\omega_2 \sigmabar^2}\big( \psi_A-\psi_V \big)  \\
{\bar{ \cal R}}^{3pt(T_3)}_3(\sigma,\omega_1,\omega_2)&=&-\frac{8\sigma^2 (m_B \sigma-\omega_1) }{\omega_2^2 \sigmabar^4} \Big\{m_B \omega_2\sigmabar \big({\bar Y}_A+{\bar W}\big)
+3(m_B \sigma-\omega_1){\bar {\bar W}} \Big\} \qquad \nn \\ \\
{\bar{ \cal R}}^{3pt(T_3)}_4(\sigma,\omega_1,\omega_2)&=&-\frac{24\sigma (m_B \sigma-\omega_1)^2 }{\omega_2^2 \sigmabar^5} \Big[\sigma (m_B^2 \sigmabar^2-q^2)+m_c^2(2-\sigma) \Big]{\bar {\bar W}} 
\eea
\end{itemize}
\newpage
\section{Light-cone distribution amplitudes of $B_q$ meson }\label{appC}
For the sake of completeness, we collect expressions of  the LCDA appearing in eqs.~\eqref{2pt1}-\eqref{3pt1}.
The functions $\phi_+,\,\phi_-,\,g_+,\,g_-$  in eqs.~\eqref{2pt1}-\eqref{barred}  have increasing twist from 2 to 5.
The exponential model proposed in \cite{Braun:2017liq} is adopted:
\bea
\phi_{+}(\omega) &=& \frac{\omega}{\lambda_B^2} \, e^{-\omega/\lambda_B}  \\
g_{+}(\omega) &=&-\frac{\lambda_E^{2}}{6\,\lambda_B^{2}}
\left[(\omega - 2\lambda_B) \operatorname{Ei}\!\left(-\frac{\omega}{\lambda_B}\right)+ \left((\omega + 2\lambda_B)
\left( \ln\!\frac{\omega}{\lambda_B}+ \gamma_E \right) - 2\omega \right) e^{-\omega/\lambda_B} \right]  \qquad \nn \\
&+& \frac{\omega^{2}}{2\,\lambda_B}
\left(1 - \frac{\lambda_E^{2} - \lambda_H^{2}} {36\,\lambda_B^{2}} \right) e^{-\omega/\lambda_B}  \\
\phi_{-}(\omega) &=& \frac{1}{\lambda_B}\, e^{-\omega/\lambda_B} - \frac{\lambda_E^{2} - \lambda_H^{2}}{9\,\lambda_B^{3}}
\left(1 - \frac{2\omega}{\lambda_B} + \frac{\omega^{2}}{2\,\lambda_B^{2}} \right) e^{-\omega/\lambda_B}\,\,.
\eea
 Although the LCDA $g_-$ is formally of twist-five, in the present framework it is fully determined by Wandzura–Wilczek–type relations and does not introduce additional independent non-perturbative parameters. Its contribution is therefore consistent with the adopted twist-four accuracy in the light-cone power expansion.
For $g_-$ the Wandzura-Wilczek approximation  gives
\be
g_{-}(\omega) = \frac{3}{4}\,\omega\, e^{-\omega/\lambda_B} \,\,\,.
\ee
 Analogous considerations hold for the three particle sector. The contribution of higher-twist three-particle LCDAs, including $\phi_6$, is included consistently within the equations-of-motion framework. Although these terms are formally of higher twist, they are power suppressed and enter only through relations among LCDAs, without affecting the truncation accuracy of the sum rule.

The functions in eq.~\eqref{3pt1} are written as
\bea
\psi_A(\omega_1,\omega_2)&=&\frac{1}{2} \left[ \phi_3(\omega_1,\omega_2) + \phi_4(\omega_1,\omega_2)\right]  \\
\psi_V(\omega_1,\omega_2)&=&\frac{1}{2}\left[-\phi_3(\omega_1,\omega_2)+\phi_4(\omega_1,\omega_2)\right]  \\
X_A(\omega_1,\omega_2)&=&\frac{1}{2}\left[-\phi_3(\omega_1,\omega_2)-\phi_4(\omega_1,\omega_2)+2\,\psi_4(\omega_1,\omega_2)\right]  \\
Y_A(\omega_1,\omega_2)&=&\frac{1}{2}\left[-\phi_3(\omega_1,\omega_2)-\phi_4(\omega_1,\omega_2)+\psi_4(\omega_1,\omega_2)-\psi_5(\omega_1,\omega_2)\right]  \\
\widetilde{X}_A(\omega_1,\omega_2)&=&\frac{1}{2}\left[-\phi_3(\omega_1,\omega_2)+\phi_4(\omega_1,\omega_2)-2\,\widetilde{\psi}_4(\omega_1,\omega_2)\right]  \\
\widetilde{Y}_A(\omega_1,\omega_2)&=&\frac{1}{2}\left[-\phi_3(\omega_1,\omega_2)+\phi_4(\omega_1,\omega_2)-\widetilde{\psi}_4(\omega_1,\omega_2)+\widetilde{\psi}_5(\omega_1,\omega_2)\right]  \\
W(\omega_1,\omega_2)&=&\frac{1}{2} \Big[\phi_4(\omega_1,\omega_2)-\psi_4(\omega_1,\omega_2)-\widetilde{\psi}_4(\omega_1,\omega_2)\nn +\widetilde{\phi}_5(\omega_1,\omega_2)+\psi_5(\omega_1,\omega_2)
\nn \\
&+&\widetilde{\psi}_5(\omega_1,\omega_2) \Big]   \\
Z(\omega_1,\omega_2)&=&
\frac{1}{4}\Big[-\phi_3(\omega_1,\omega_2)+\phi_4(\omega_1,\omega_2)-2\,\widetilde{\psi}_4(\omega_1,\omega_2)+\widetilde{\phi}_5(\omega_1,\omega_2)
+2\,\widetilde{\psi}_5(\omega_1,\omega_2)\nn \\
&-&\phi_6(\omega_1,\omega_2) \Big]  ,
\eea
where \cite{Braun:2017liq}  
\bea
\phi_3(\omega_1,\omega_2) &=&
\frac{\lambda_E^{2}-\lambda_H^{2}}{6\,\lambda_B^{5}}\,\omega_1 \omega_2^{2}\,e^{-(\omega_1+\omega_2)/\lambda_B} \\
\phi_4(\omega_1,\omega_2)
&=&\frac{\lambda_E^{2}+\lambda_H^{2}}{6\,\lambda_B^{4}}\,\omega_2^{2}\, e^{-(\omega_1+\omega_2)/\lambda_B} \\
\psi_4(\omega_1,\omega_2)
&=&\frac{\lambda_E^{2}}{3\,\lambda_B^{4}}\,\omega_1 \omega_2\, e^{-(\omega_1+\omega_2)/\lambda_B}  \\
\widetilde{\psi}_4(\omega_1,\omega_2)
&=& \frac{\lambda_H^{2}}{3\,\lambda_B^{4}}\,\omega_1 \omega_2\, e^{-(\omega_1+\omega_2)/\lambda_B} 
\eea
and  \cite{Lu:2018cfc} 
\bea
\psi_5(\omega_1,\omega_2)
&=&-\frac{\lambda_E^{2}}{3\,\lambda_B^{3}}\,\omega_2\, e^{-(\omega_1+\omega_2)/\lambda_B}
 \\
\widetilde{\psi}_5(\omega_1,\omega_2)
&=&-\frac{\lambda_H^{2}}{3\,\lambda_B^{3}}\,\omega_2\, e^{-(\omega_1+\omega_2)/\lambda_B}
 \\
\widetilde{\phi}_5(\omega_1,\omega_2)
&=&\frac{\lambda_E^{2}+\lambda_H^{2}}{3\,\lambda_B^{3}}\,\omega_1\, e^{-(\omega_1+\omega_2)/\lambda_B}
 \\
\phi_6(\omega_1,\omega_2)
&=&\frac{\lambda_E^{2}-\lambda_H^{2}}{3\,\lambda_B^{2}}\, e^{-(\omega_1+\omega_2)/\lambda_B} \,\,.
\eea
The parameters $\lambda_B$,  $\lambda_E$, and $\lambda_H$ have been determined for $B_d$   \cite{Braun:2003wx,Nishikawa:2011qk} and  $B_s$   \cite{Mandal:2024pwz}. The numerical values 
are collected in table~\ref{tab:lambdas}.

\begin{table}[h]
\centering
\renewcommand{\arraystretch}{1.3}
\begin{tabular}{lcc}
\hline \hline
 & $B_d$   \cite{Braun:2003wx,Nishikawa:2011qk} & $B_s$  \cite{Mandal:2024pwz} \\
 \hline 
 $\lambda_B$& $0.460 \pm 0.110 \,\,\,  {\rm GeV}$ & $0.438 \pm 0.150 \,\,\, {\rm GeV}$\\
 $\lambda_E^2$& $0.02 \pm 0.03 \,\,\, {\rm GeV}^2$ &  $0.03 \pm 0.02 \,\,\, {\rm GeV}^2$\\
 $\lambda_H^2$& $0.11 \pm 0.08 \,\,\, {\rm GeV}^2$& $0.06 \pm 0.03 \,\,\,{\rm GeV}^2$  \\
\hline \hline
\end{tabular}
\caption{\baselineskip 10pt \small Values of the parameters $\lambda_B$,  $\lambda_E$ and  $\lambda_H$   used  for the $B_d$  and   $B_s$ form factor computation.}\label{tab:lambdas}
\end{table}

\newpage
\bibliographystyle{JHEP}
\bibliography{refCFGP}
\end{document}